\documentclass[12pt,preprint]{aastex}

\usepackage{graphicx}

\newcommand{\arcs}{\ensuremath{^{\prime\prime}}}

\newcommand{\kms}{km\ ${\rm s^{-1}}$}
\newcommand{\ms}{m\ ${\rm s^{-1}}$}
\newcommand{\mxcm}{Mx cm$^{-2}$}
\shorttitle{Explosive Events and Magnetic Fields}
\shortauthors{Muglach}

\begin{document}

%% ----------------------------------------------------------------------
%% --- title page
%% ----------------------------------------------------------------------

\title{Explosive Events and the Evolution of the 
       Photospheric Magnetic Field}

\author{K.~Muglach}
\affil{Naval Research Laboratory, Washington, DC 20375, USA;
       }
 %       \\email: muglach at nrl.navy.mil}
\altaffiltext{1}{Also at Artep Inc., Ellicott City, MD 21042, USA}

%% ----------------------------------------------------------------------

\begin{abstract}
Transition region explosive events have long been 
suggested as direct signatures of magnetic reconnection in
the solar atmosphere. In seeking further observational evidence to 
support this interpretation, we study the relation between explosive
events and the evolution of the solar magnetic field as seen
in line--of--sight photospheric magnetograms.
We find that about 38\% of events show changes of the magnetic
structure in the photosphere at the location of an explosive
event over a time period of 1 h.
We also discuss potential ambiguities in the analysis of high
sensitivity magnetograms.
\end{abstract}

 
%----------------------------------------------------------------------

\section{Introduction \label{sec:intro} }

Explosive events (EEs) are defined by strong transient enhancements
of the wings of spectral lines that form at transition
region (TR) temperatures. They were first observed with the
rocket--borne High Resolution Telescope and Spectrograph (HRTS)
flown by the Naval Research Laboratory
(Brueckner \& Bartoe 1983; Dere et al.~1984).
The physical properties of EEs as derived from the HRTS flights
are described in Dere (1994, Table 1): time scales between
20-200s (with an average of 60s), spatial scales of around
1500 km ($\approx$2\arcs ) and velocities of about
100 \kms . They are best seen in TR emission lines produced at
temperatures of $2 \times 10^4 - 2 \times 10^5$ K.
Spatial maps derived from raster sequences showed that
EEs can be found all over the quiet Sun, including at the limb
and in coronal holes.
Wing enhancements were not always symmetric, sometimes
only a red or blue wing was present, often having a spatial
offset of 1\arcs\--2\arcs\ along the slit.
%It was also noted that there is hardly any increase in
%intensity of the central line component.

More recent studies have analysed spectra from the 
Solar Ultraviolet Measurements of Emitted Radiation
(SUMER, Wilhelm et al.~1995) instrument onboard 
the Solar and Heliospheric Observatory (SoHO).
P\'erez \& Doyle (2000) summarize the properties
of EEs derived from SUMER spectra,
based on a small sample of 8 events.
Earlier HRTS results are largely confirmed, with some
variations due to the different characteristics of the
SUMER instrument (see also Teriaca et al.~2004).
For instance, lifetimes are longer, probably due to the longer
exposure times needed by SUMER, while sizes are larger 
(4\arcs--14\arcs ) due to the lower spatial resolution of
SUMER compared to HRTS.

Part of the interest in the study of EEs stems from the idea
that they could be an important indicator of heating and/or
acceleration processes in the upper solar atmosphere.
To determine the origin of EEs and evaluate their contribution 
to the mass and energy flux in the solar atmosphere,
one must address the question of their relationship with the
magnetic field topology.
Porter \& Dere (1991) found that most EEs are located
near the magnetic network, which was confirmed later, e.g.~by Moses
et al.~(1994) and Chae et al.~(1998).
The TR line emission of EEs is usually fainter than the enhanced
emission that corresponds to the magnetic network.
This is one of the important differences between EEs and
other small--scale transients of the solar TR and corona.
For example, coronal and X-ray bright points are found directly
within the bright network and exhibit enhanced X-ray and EUV emission,
with sizes of about 10\arcs\ -- 40\arcs\ and lifetimes between several
hours and 1--2 days. They consist of small--scale arcades of
loops connecting opposite polarities of small magnetic
bipoles (e.g.~Sheeley \& Golub 1979; Habbal 1992; Webb et al.~1993;
Madjarska et al.~2003).
In contrast, EEs are usually not seen in coronal lines, but
are limited to lines at TR temperatures.

Another TR phenomenon which was discovered with 
Coronal Diagnostic Spectrometer (CDS) on SoHO
are the so-called blinkers (e.g.~Harrison 1997, Bewsher et al.
2002). They are best observed in transition region lines like
O IV 554 \AA\ ($1.6\times 10^5$ K)
and O V 629 \AA\ ($2.5\times 10^5$ K).
These events are characterized by an intensity
enhancement factor of about 1.2 to 2.7 and a duration
of around 1 to 40 min (although some of the longer ones
seem to be composed of short--duration elementary blinkers,
according to Brooks et al.~2004). Like EEs, blinkers do not
appear to have any detectable coronal or chromospheric signatures.
Bewsher et al. (2002) found that most blinkers are
located above regions of strong flux, the magnetic network.

The relationship between EEs and TR blinkers has been examined
in several studies. Based on a statistical analysis Brkovi{\'c}
\&  Peter (2004) concluded that explosive events and blinkers are
independent phenomena. Bewsher et al.~(2005) raise doubts as to
a link between them and their probability analysis shows that
such a link is very unlikely.
Madjarska \& Doyle (2003) described the
main characteristics of both EEs and blinkers 
that imply that they are more likely two separate
phenomena not directly related or triggering each other.
On the other hand, Chae et al.~(2000) 
have put forth the idea that blinkers ``consist of many
small--scale and short--lived SUMER unit brightening events
having sizes and durations that are very comparable to those
of explosive events''. They attribute the difference in
spectral characteristics of blinkers and EEs to different
magnetic geometries involved in the underlying reconnection.
This controversy clearly underscores the importance of 
studying the magnetic field structure of EEs and other
small-scale TR phenomena.

The basic mechanism proposed for the formation of
EEs is magnetic reconnection taking place near the
magnetic network. Dere et al.~(1991,
their Fig.~2) illustrates this in a schematic way.
Patches of bipolar magnetic flux emerge in the
interiors of the supergranular cells and
are swept to the boundaries by the supergranular
flow. If a network element has the opposite
polarity, the flux will eventually come close enough
for reconnection to occur.
The intense wing broadening of the spectral lines is
thought to be due to plasma that is accelerated in
the reconnection process (Innes et al.~1997).
In this scenario the EEs are therefore considered to
be a direct consequence of magnetic reconnection in the TR.

From the spatial distribution of EEs superposed on a map
that displays the chromospheric network
it can be seen that EEs not only seem to avoid the
magnetic network, but are also rarely located in the
centers of supergranular cells (e.g. Porter \& Dere 1991).
The work of Porter \& Dere (1991) was limited to a comparison of
EE locations with single magnetograms taken at the time
of the HRTS flights. Therefore, the temporal evolution of the
magnetic field could not be addressed.

Some of the early observations with HRTS showed some
associations of EEs with emerging flux regions, as in the case
of HRTS-6 (Dere et al. 1991).
For the HRTS-7 flight magnetograms from several ground--based
observatories were available. In one case an EE was found to
be connected to a small emerging bipole that later merged with
a pre-existing patch of nearby flux (Dere \& Martin 1994).
Many more EEs were found in quiet sun regions, but no
simultaneous magnetic field measurements were available. Therefore, the
case for interpreting EEs as evidence for reconnection during
magnetic flux emergence has been largely circumstantial.

A detailed account of the magnetic field evolution
in connection with EEs is given by Chae et al.~(1998).
They compared SUMER EEs with deep magnetograms taken
at Big Bear Solar Observatory (BBSO). They
found that EEs are located near the magnetic network
(in agreement with Porter \& Dere 1991).
Out of a total of 163 EEs in their spectroheliogram, 103
are at a location of flux cancellation.
Section \ref{sec:disc} discusses their results in more detail.

More recently, Madjarska \& Doyle (2003) used data from SUMER,
CDS and BBSO and were unable to associate EEs with any particular
magnetic field pattern. On the other hand they found that
blinkers are associated with bipolar magnetic regions, although
they found no flux cancellation or changes in total flux for
blinkers.

A different approach to a closely related topic was presented by
Sanchez Almeida et al. (2007). Their work illustrated an attempt
to find footpoints of quiet Sun TR loops using G--band bright points.
These are used as proxies for photospheric magnetic field
concentrations which may anchor and guide the TR structures.
They also identified several EEs in their SUMER data, but
found a tendency that EEs avoided G--band bright points.
As explained in our standard scenario above, we consider EEs are
the direct signature of reconnection taking place in the TR.
On the other hand Tarbell et al.~(2000) proposed reconnection in the 
photosphere, which leads to shocks and shock interaction
that accelerate the plasma at TR temperatures.
Sanchez Almeida et al.~(2007) argued that, if G--band bright
points are footpoints of loops undergoing reconnection, then the
fact that the EEs do not coincide with them indicates that the site
of TR plasma acceleration is far from the photospheric footpoints.

%Another point of view concerning the site of the reconnection
%within the solar atmosphere was given by Benz \& Krucker (1999).
%They assume reconnection high in the corona
%which generates high energy particle beams that heat and
%accelerate the chromospheric plasma leading to the TR signature.

The purpose of the current study is to
identify signatures of EEs in the underlying
photospheric magnetic field and to illuminate
the role of magnetic reconnection in the formation of EEs.
The following section describes the observations
and basic data analysis. Section \ref{sec:res}
shows the results of our combined data and
Section \ref{sec:disc} discusses the magnetic field evolution.

%__________________________________________________________________

\section{Observations and Data Processing \label{sec:obs} }

In our analysis we combine data from two instruments onboard
SoHO. Until recently the SUMER instrument was the
only instrument in operation that was able to identify EEs
by their spectral signature. The Michelson Doppler Imager (MDI,
Scherrer et al.~1995) can continuously map a large disk fraction of
photospheric magnetic field with moderate to good spatial
resolution over many hours. The data have been taken on 7th
November 1999, they have been taken strictly simultaneously
and cospatiality is ensured by the data processing.
With this combination of instruments we have the
opportunity to study the relation of EEs and the
underlying magnetic field in unprecedented detail.

\subsection{SUMER \label{sec:obssum} }

%SUMER data from 5.11.99 have also been used by Xia, Marsch and
%Curdt (2003).

\begin{figure}[htbp!]
\centering
\vbox{
\hbox{
    \includegraphics[width=0.5\textwidth]{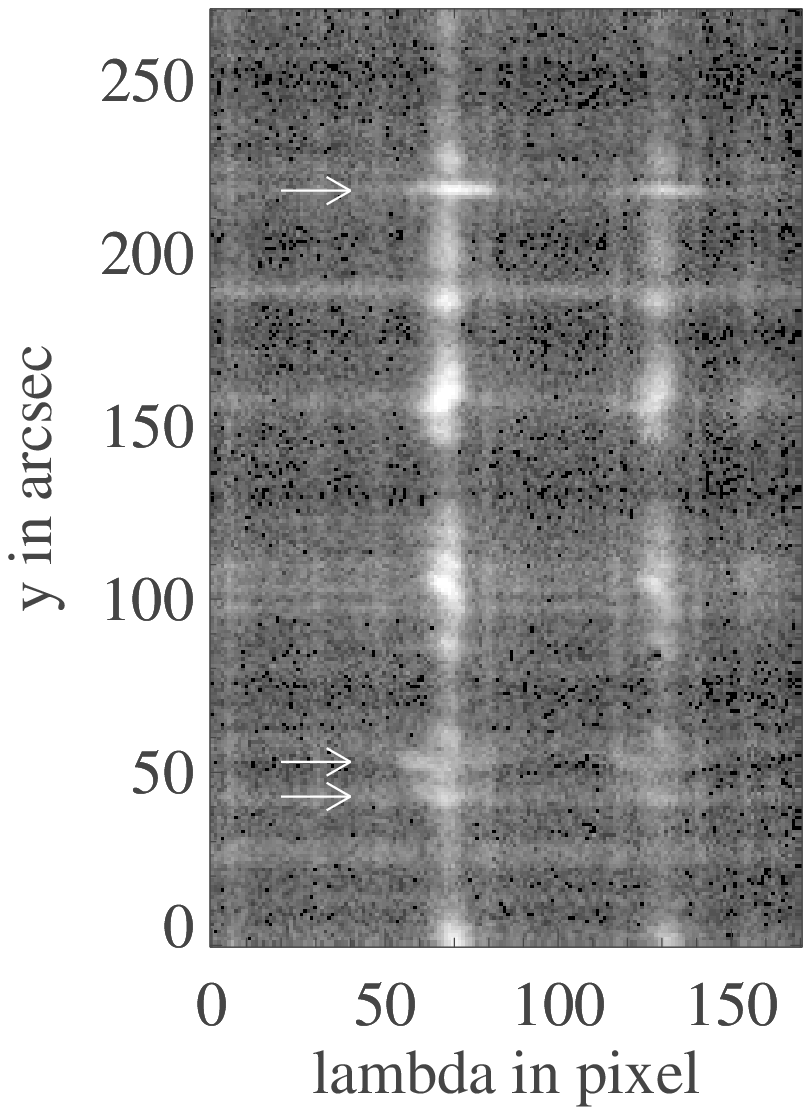}
    \includegraphics[width=0.5\textwidth]{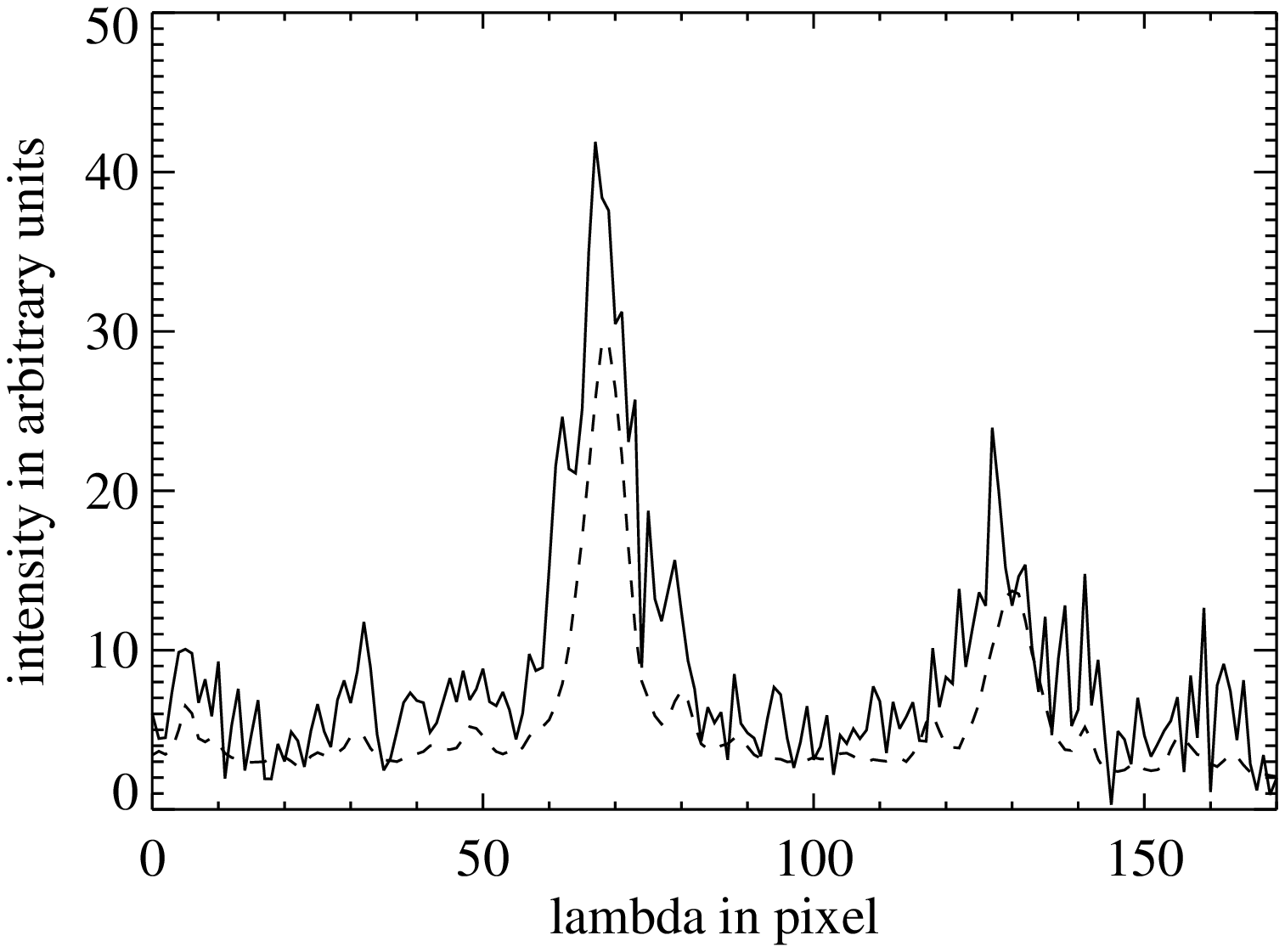}
     }
\hbox{
    \includegraphics[width=0.5\textwidth]{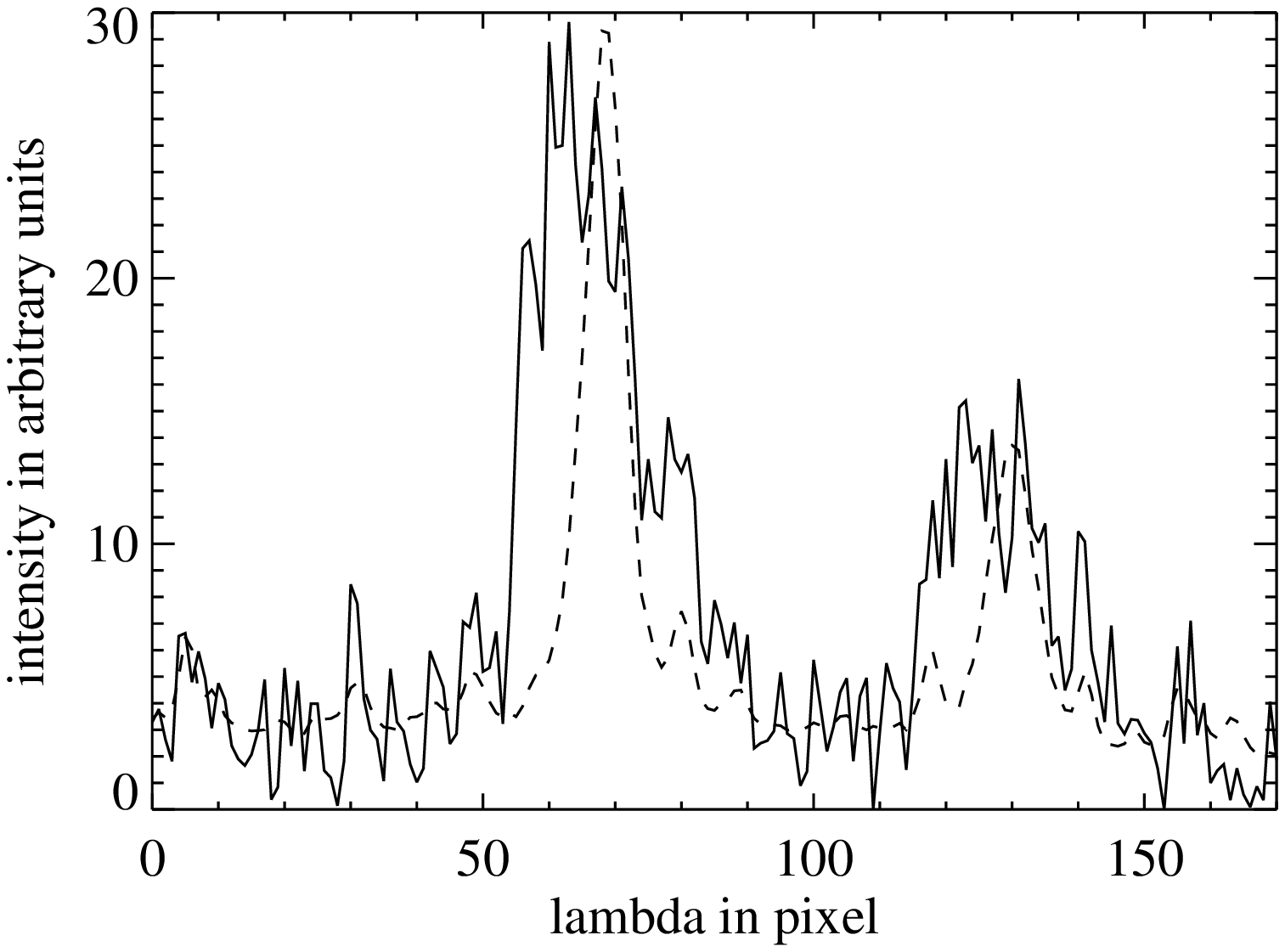}
    \includegraphics[width=0.5\textwidth]{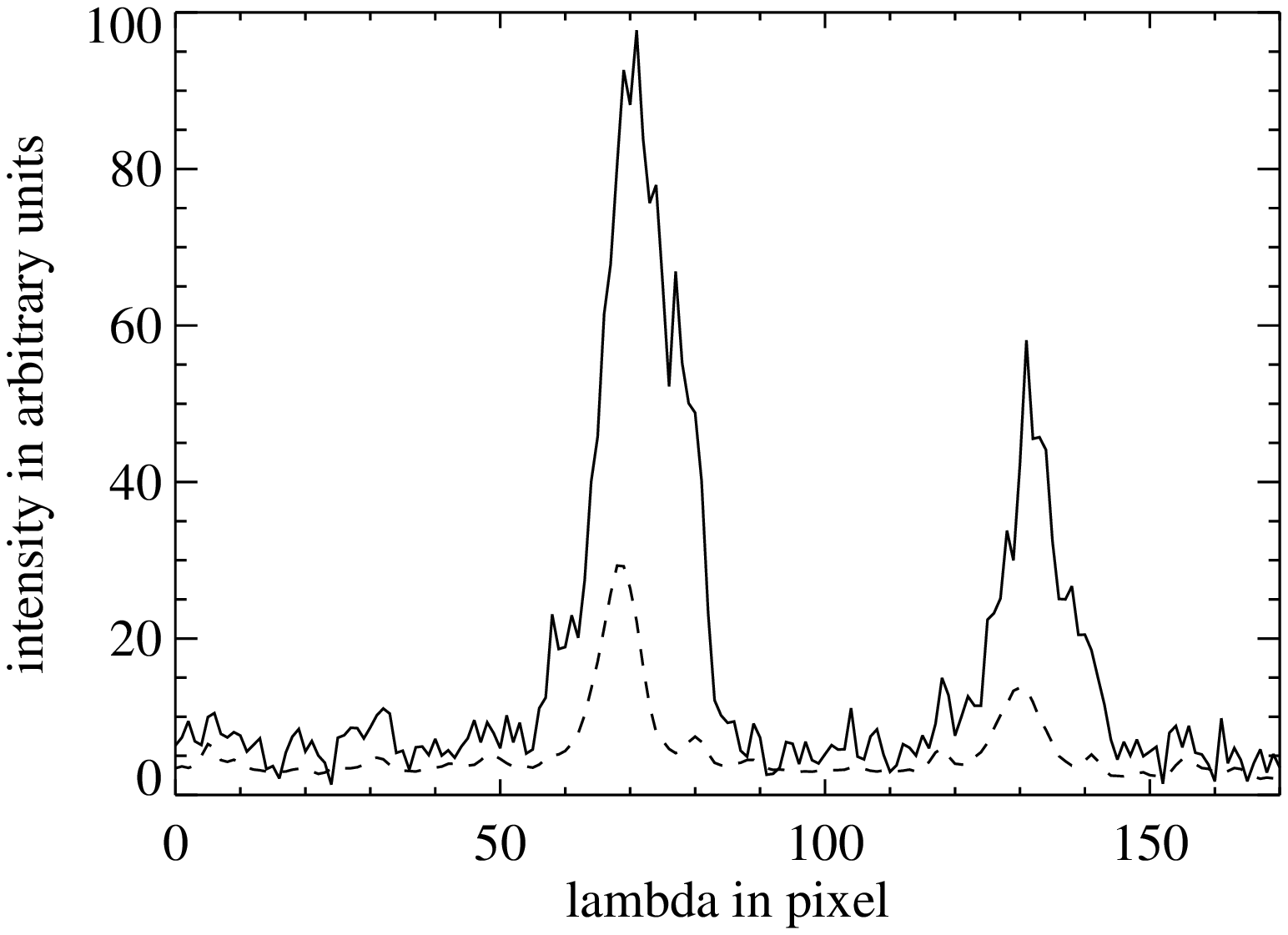}
   }
}
\caption{
        Example of C IV spectra at 1548 \AA\ and 1550 \AA\
        (left), the white arrows point
        out the location of the EEs in this particular
        slit position. The line plots show the line profiles
        at the indicated arrow positions, the solid line represents
        the EE profile, the dashed line a profile averaged along
        the slit.
        }
\label{fig:SUMER-EE-examp}
\end{figure}

The SUMER instrument onboard SoHO was designed to observe 
spectral lines in the range from around 500 \AA\ to
1610 \AA . This study employs data obtained
in the scanning mode in order to allow coalignment of
the SUMER data with the MDI magnetograms.
SUMER was taking data around 1550 \AA ,
which covered two C IV lines (1548 \AA , and 1550 \AA ,
formation temperature $1\times 10^5$ K), a Si II line (1533 \AA ,
$1.8\times10^4$ K) and a Ne VIII line (770 \AA\ in the second order,
$6.3\times10^5$ K).
Figure \ref{fig:SUMER-EE-examp} shows an example of a C IV
spectrum, including several spectral profiles of EEs.
Exposure times were 150 s and SUMER performed 60 raster steps
each of 3\arcs\ (nominally). This scan was repeated 3 times.
Only the first scan and part of the second had
co-temporal MDI data and are considered in this analysis.
Spatial sampling along the SUMER slit corresponds to
approximately 1\arcs /pixel.

Basic data processing of the raw SUMER spectra includes
decompression, flat--fielding and geometrical corrections.
We note that the destretching algorithm 
that is available with the public software uses pre-flight
calibration data and does not work perfectly.
Residual image distortions are still present and are
worst at the edges of the detector. 
Finally, we fit the line profiles with a single Gau\ss ian 
to obtain integrated intensity, central wavelength and width
of the spectral lines.

The characteristic signature of an EE is enhanced
emission in the wings of the spectral line. We developed
the following algorithm to identify such wing enhancements.
We first determine the continuum in a nearby wavelength
range and the root mean square (rms) of the continuum
$I_{\rm rms}$.
We then smooth the spectra in wavelength and determine the
location of the maximum emission in the C IV 1548 \AA\
line. We exclude locations where the maximum of the
emission is less than 10$\times I_{\rm rms}$ and where the
continuum is less than 1. Thus, the algorithm avoids
very weak profiles
that mostly consist of noise. We note that these
criteria automatically remove most locations in the
centers of network cells.
Therefore our algorithm is biased against
finding EEs in cell centers.
%Not finding EEs in cell centers
%is in our case a bias of our algorithm.
Finally, we calculate the average intensity between
70--130 \kms (5 spectral pixels) to the red and blue of
the line center.
If this value is larger than 0.3 times the maximum intensity
of the line, the line profile will be flagged as a potential
EE. In the last step, the flagged spectra are displayed
and the final determination is done visually.

Our algorithm is designed to reject moderately enhanced
(less than 30\% of the line maximum)
and moderately broadened (velocities $<$ 70 \kms) line profiles.
While such profiles exist, they are generally not included
in our analysis. In the case of rather weak lines the 30\%
criterion removes profiles where the wing enhancements are
also near the noise level. On the other hand, we found several
cases where the central component of the line was very strong
and a very clear wing enhancement was present, but the
30\% criterion was not fulfilled due to the brightness
of the central component.  We have added these locations
to our sample of EEs by hand.

In agreement with the known size of EEs,
characteristic wing enhancements are often
found in more than one pixel along the
slit and sometimes also in neighbouring scan
positions. If an event covers several pixels,
all adjacent pixels are counted as one single event.

\subsection{MDI \label{sec:obsmdi} }

We use magnetograms and Dopplergrams that were taken
in the high resolution (HR) mode of MDI.
These data have a spatial sampling of
0.6\arcs\ per pixel, and one set of observables
is obtained every minute.

The data used in this study are available from the MDI web
archive. and have been processed to the 1.8 level.
Raw data from the spacecraft have first been calibrated into physical
units of the observable and time and location of the observation.
Descriptive and ancillary data have been added and finally
level 1.8 processing went on to provide time--dependent corrections
for plate scale, zero offset, sensitivity, and cosmic rays.
We checked the sequences for missing
images and used linear interpolation from adjacent
ones to fill data gaps.

It takes 30 s to measure the 5 filter locations in
the Ni I 6768 \AA\ line from which magnetograms,
Dopplergrams and intensity images are derived.
According to Scherrer et al. (1995) a single longitudinal
magnetogram has a 1$\sigma$ noise level of 20 \mxcm .
At least part of the noise in the magnetograms can be 
attributed to cross-talk of the velocity
signal into the magnetic signal (e.g.~Settele et al. 2002).
During the time MDI needs to measure the 5 filter
positions (30 s), the spectral line is also shifted in 
wavelength due to the photospheric velocity
field. The magnetic signal is calculated from differences
of filtergrams in the two polarization states, and the Doppler
shift introduces such differences over the exposure time. 

We carry out a 3--D Fourier transformation of our quiet
Sun MDI magnetogram sequences. In the resulting
$k{-}\omega$ diagram variations with characteristic time scales
of 2--5 min are present which are due to wave motions.
To remove the effect of oscillations in our magnetograms,
we apply a subsonic Fourier filter
with a cutoff phase velocity of $v_{ph}$ = 3 \kms .
%Title et al.~(1989)
Close inspection of the filtered magnetogram sequences
reveals the presence of fluctuations on granular scales.
As we expect the changes in magnetic flux associated with
EEs to be small, we refrain from additional spatial and/or
temporal averaging.

To determine the noise in the filtered magnetograms
we follow the method of Hagenaar (2001) and Parnell (2002).
A histogram of the magnetic flux of various selected
regions was fit with a Gau\ss ian centered on the zero value.
From this analysis we derive a value
of $B = \pm$ 9 Mx cm$^{-2}$ as the one sigma noise level.
We disregard all flux between these values by setting it to zero.

We apply the same analysis procedure to the MDI Dopplergrams.
The rms of the Doppler fluctuations of a filtered data-cube
is around 150 \ms , with a peak--to--peak amplitude of 
about $\pm$ 600 \ms .

%__________________________________________________________________

\subsection{Combining SUMER and MDI data \label{sec:sumandmdi} }
%\section{Data processing \label{sec:proc} }

%use data4, 3 SUMER scans, 2 have MDI overlap, MDI HR data,

Because EEs have very small spatial scales, proper co--alignment
of SUMER with MDI is critical. Both the SUMER scan
and MDI magnetograms have to be scaled to the same size
and then cross-correlated. The fact that SUMER is scanning
while MDI takes a series of snapshots complicates the
analysis further due to the evolution of the solar
structures.

SUMER scanning is carried out either from east to west
or from west to east. Depending on whether or not the scan
is in the sense of the Sun's rotation, the observed
field of view (FOV) is either smaller or larger.
The smallest scan step that can be carried out with the
scanning mechanism is 0.38\arcs , so a complete scan step
of 3\arcs\ requires 8 elementary steps. After 1997
the scanning mechanism has been unreliable and
sometimes elementary steps are not carried out.
We begin by adopting the nominal 3\arcs\ stepsize and 
add/subtract the solar rotation to determine the
size of the SUMER scan.
For the co--alignment we use the continuum emission
near the C IV 1548 \AA\ line, which represents
emission from the chromosphere.
Simply overlaying the SUMER continuum map on
the MDI magnetograms is inadequate because
of inaccuracies in the SUMER destretching and the
variable step size of the SUMER scan.
Instead we apply the following two--step procedure.
First we cross--correlate the complete scan with the
magnetogram, choosing a spatial scaling
factor that gives the optimal match. Then, we select a smaller
15\arcs$\times$25\arcs\ FOV around each EE location and
calculate the cross--correlation again. For each
event the quality of the cross-correlation
of the subfields is visually checked; if spatial distortions
are clearly present, the event is rejected.

A spatial offset between the location of the EE and of the
flux changes can be due to uncertainties in
the spatial scaling (especially in the scan direction),
residual errors in the spatial cross-correlation, and
the fact that the photospheric magnetic signal
originates from a different height than the EE.
Taking all of these effects into account, we search for magnetic
flux changes in a radius of about 6\arcs (10 MDI HR pixels) around
the location of the SUMER EE.
We then study the evolution of the magnetic flux starting 30 min
before and lasting until 30 min after the time of the EE.

Figure \ref{fig:SUMER-MDI-scan} gives an overview of one
of the scans. Abscissa $x$ is along the scanning direction,
while the SUMER slit was oriented along the ordinate $y$.
The left image shows the continuum intensity around 1535 \AA .
The bright structures outline the magnetic network in an area near disk
center. A comparison with full disk EIT coronal images shows
that the scans were taken inside a coronal hole.
The central image shows the peak intensity of the
C IV 1548 \AA\ line, with the highlighted areas indicating
locations of EEs. The right image shows a magnetogram
composed of a slice through the MDI 3-D data cube, similar to
the SUMER scan.

%__________________________________________________________________

\section{Results \label{sec:res}  }

Figure \ref{fig:MDI-evol} presents several examples
of magnetic flux changes observed at the locations of EEs.
The images are saturated at
$B = \pm$30 Mx cm$^{-2}$. In each example the left image is
a magnetogram taken 30 min before the EE took place, the
central image shows the distribution
of magnetic flux at the time of the EE, which is located at
the center of the FOV (as indicated by the white
cross), and the right image is the magnetogram taken 
30 min after the EE.
Movies of these and several other events can be found at
http://wwwsolar.nrl.navy.mil/$\sim$muglach/ee.html .

The following description of the magnetic flux evolution
refers to the complete movies, where Figure \ref{fig:MDI-evol}
only shows snapshots for each case.
An example of flux cancellation can be seen in the top row
of Figure \ref{fig:MDI-evol}. Near the center of the FOV there
is a large patch of white--polarity flux with a small piece of
black--polarity flux next to it. Just to the north there is a
large black--polarity patch with a small white--polarity area
on its east side. In both places the minority polarity moves
towards the larger magnetic patch and slowly disappears.
Both areas of cancelling flux lie within our 6\arcs\ radius
around the EE location.

The second row of Figure \ref{fig:MDI-evol} shows another example
of flux cancellation: again, a very small patch of one
(black) polarity approaches a larger patch of the other
(white) polarity and completely vanishes by the end of
the 1h sequence. 

The third row Figure \ref{fig:MDI-evol} illustrates an example
of flux emergence at the EE site. Although there is a bipole present
near the center of the FOV at the beginning of the sequence, a
second small bipole appears subsequently. The new bipole eventually
becomes clearly visible in the right image at the end of the 1h period.

The fourth row shows complex flux changes: during the first
half an hour a small bipole emerges, which cancels again in
the second half an hour. 
Finally the bottom row is a typical example where we do not
find any flux changes around the EE site.

Table \ref{tbl-1} summarizes our results. Out of
a total of 37 EEs we find that 14 (38\%) show flux changes in
a time interval of 1h around the time of the EE.
We have 3 cases of flux emergence, 7 cases of flux 
cancellation and 4 complex cases with both
emergence and cancellation taking place
within our 6\arcs\ radius of the EE location.
For 23 EEs (62\%) we could not identify any significant
flux changes.
In the second line of Table \ref{tbl-1} we show the distribution
of flux changes at randomly selected locations
in the SUMER scans. Only 4 (11\%) of the random locations
show flux changes similar to the ones we find at EE locations.

%Total EE:  37: 14(yes)/23(no)  38% / 62%
%total random: 37: 4(yes)/33(no)   11% / 89 %
%   emergence      cancellation     both/complex     none
%       3               7                  4          23
%       2               1                  1          33

%check: find papers that give statistics of small-scale flux emergence
%and flux cancellation - hagenaar? S. Martin???

Visual inspection of the MDI Dopplergram sequences did not reveal any
signatures we might expect in connection with the restructuring of
the magnetic field.

%__________________________________________________________________

\section{Discussion \label{sec:disc} }

\begin{figure}[htbp!]
\centering
\hbox{
    \hspace{-3cm}
    \includegraphics[width=0.65\textwidth]{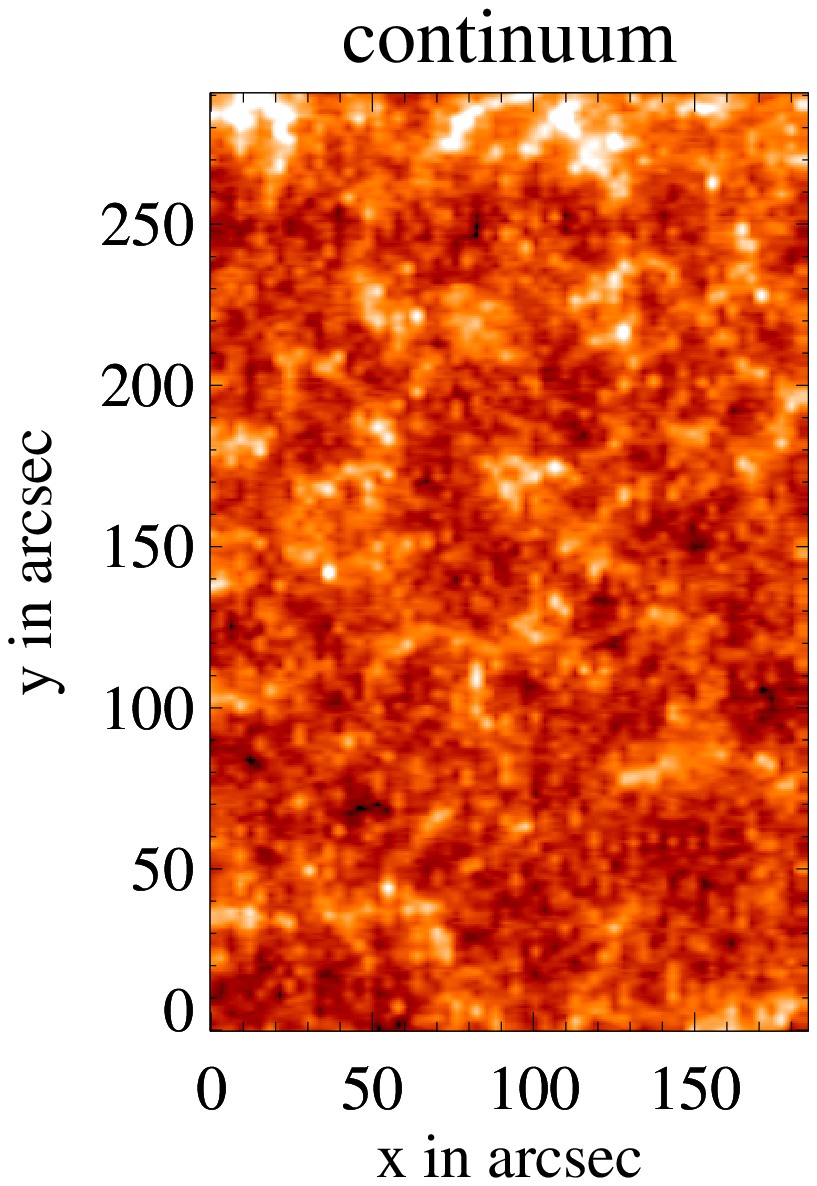}
    \hspace{-6cm}
    \includegraphics[width=0.65\textwidth]{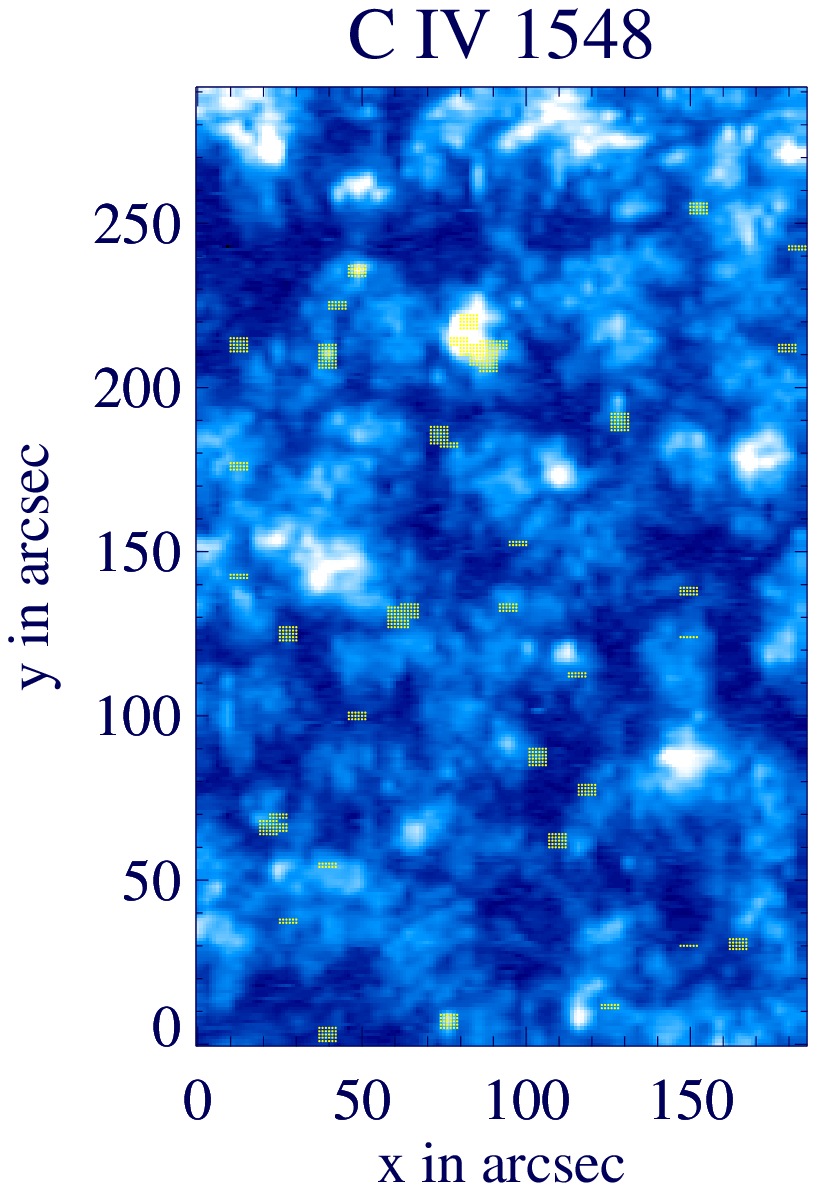}
    \hspace{-6cm}
    \includegraphics[width=0.65\textwidth]{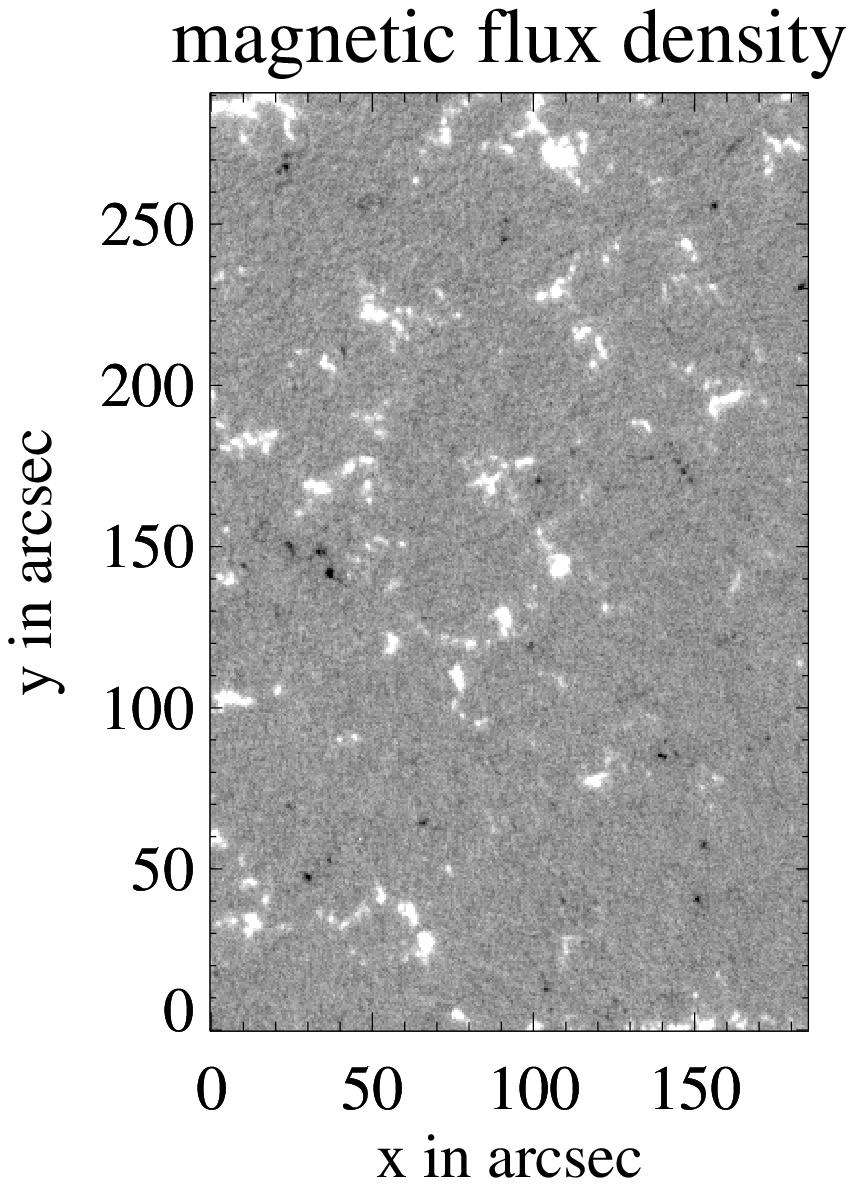}
}
\caption{Example of a SUMER scan. Left: image of 
continuum intensity around 1535 \AA .
Center: the peak intensity of C IV 1548 \AA .
The yellow areas are locations of EEs,
identified by strong wing enhancements in the C IV line.
Right: magnetogram composed of a slice through the MDI
3-D data cube, following the SUMER scan.
        }
\label{fig:SUMER-MDI-scan}
\end{figure}

\begin{figure}[htbp!]
\centering
\vbox{
\hbox{
      % \hspace{-1cm}
    \includegraphics[width=0.35\textwidth]{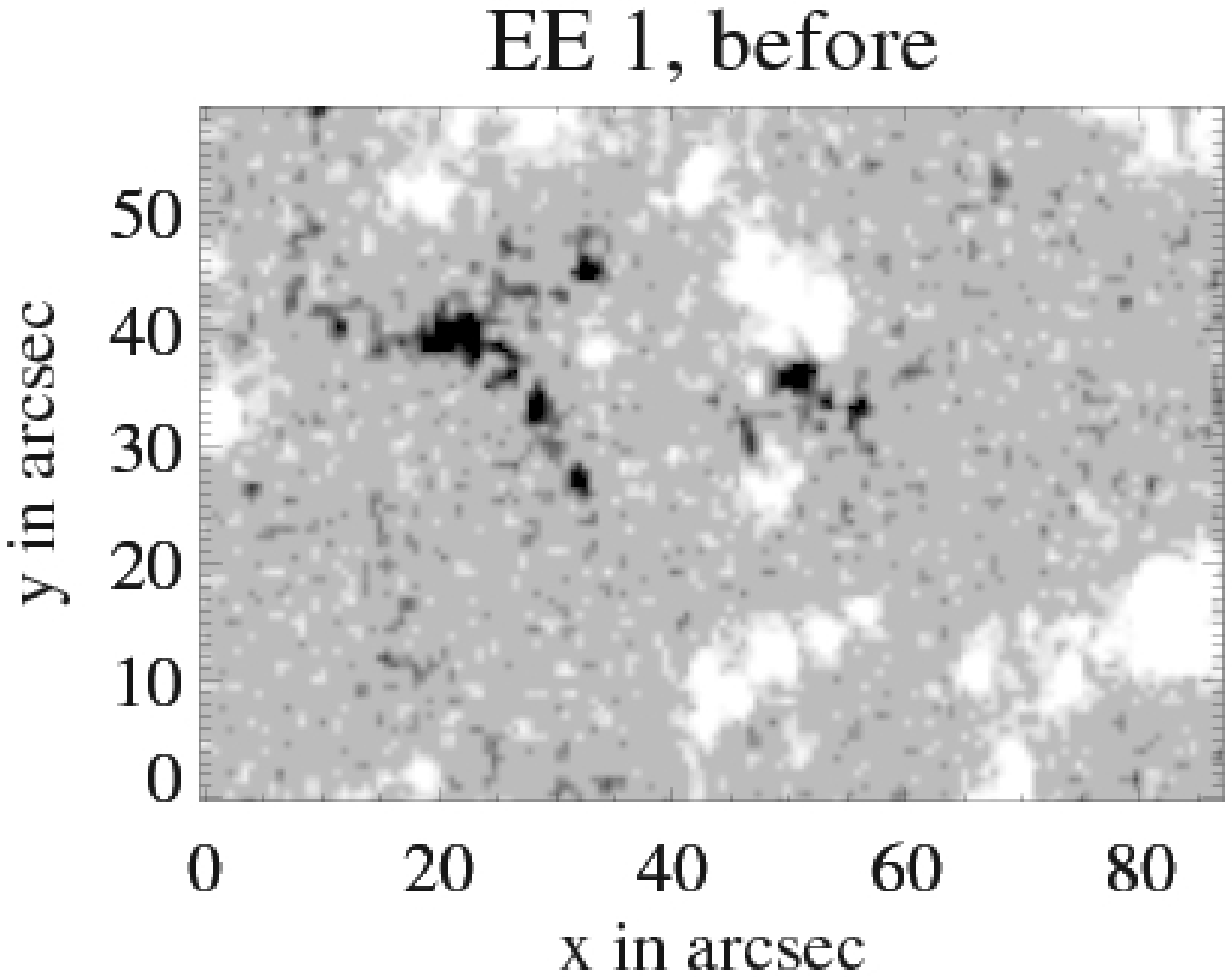}
       \hspace{-1cm}
    \includegraphics[width=0.35\textwidth]{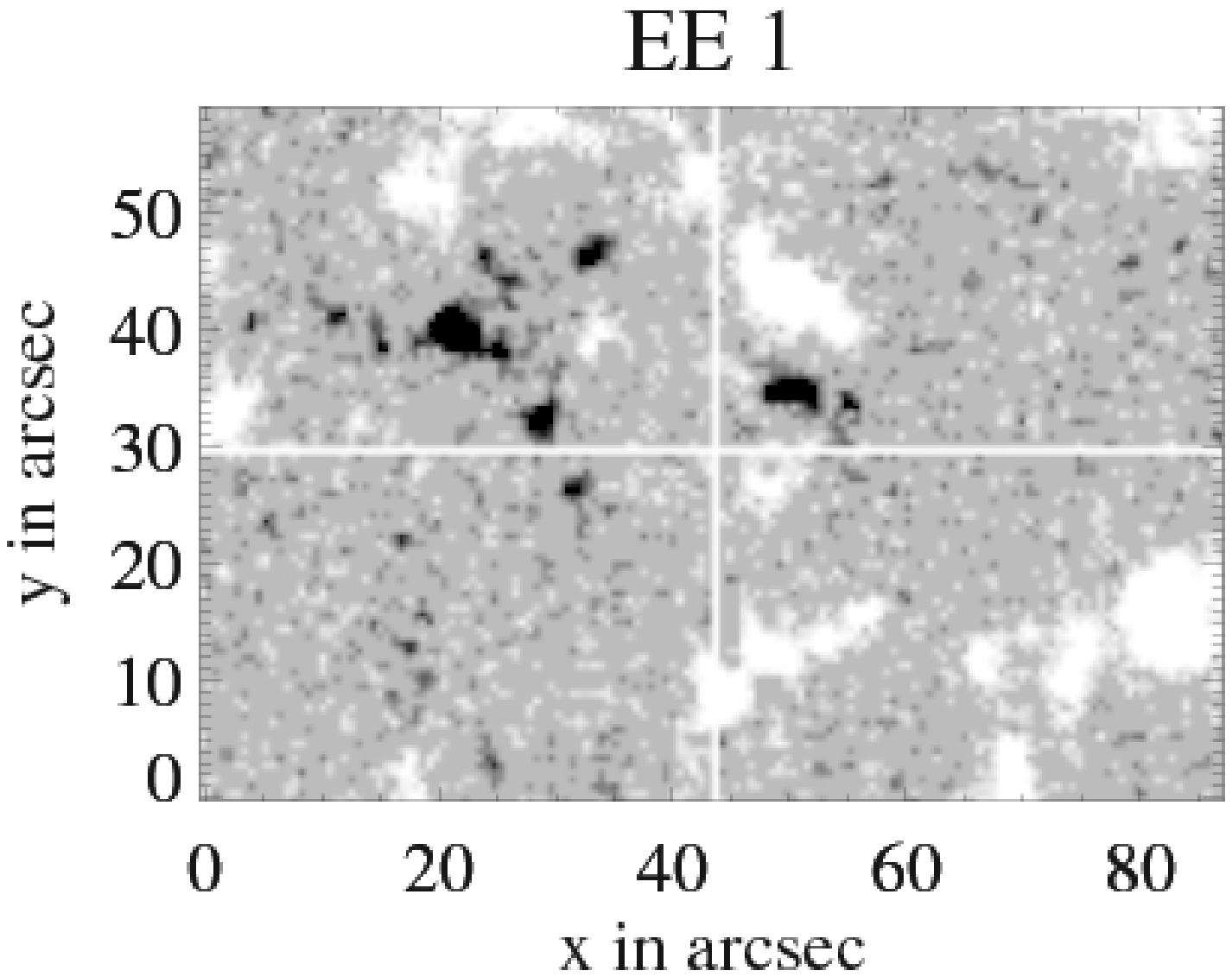}
       \hspace{-1cm}
    \includegraphics[width=0.35\textwidth]{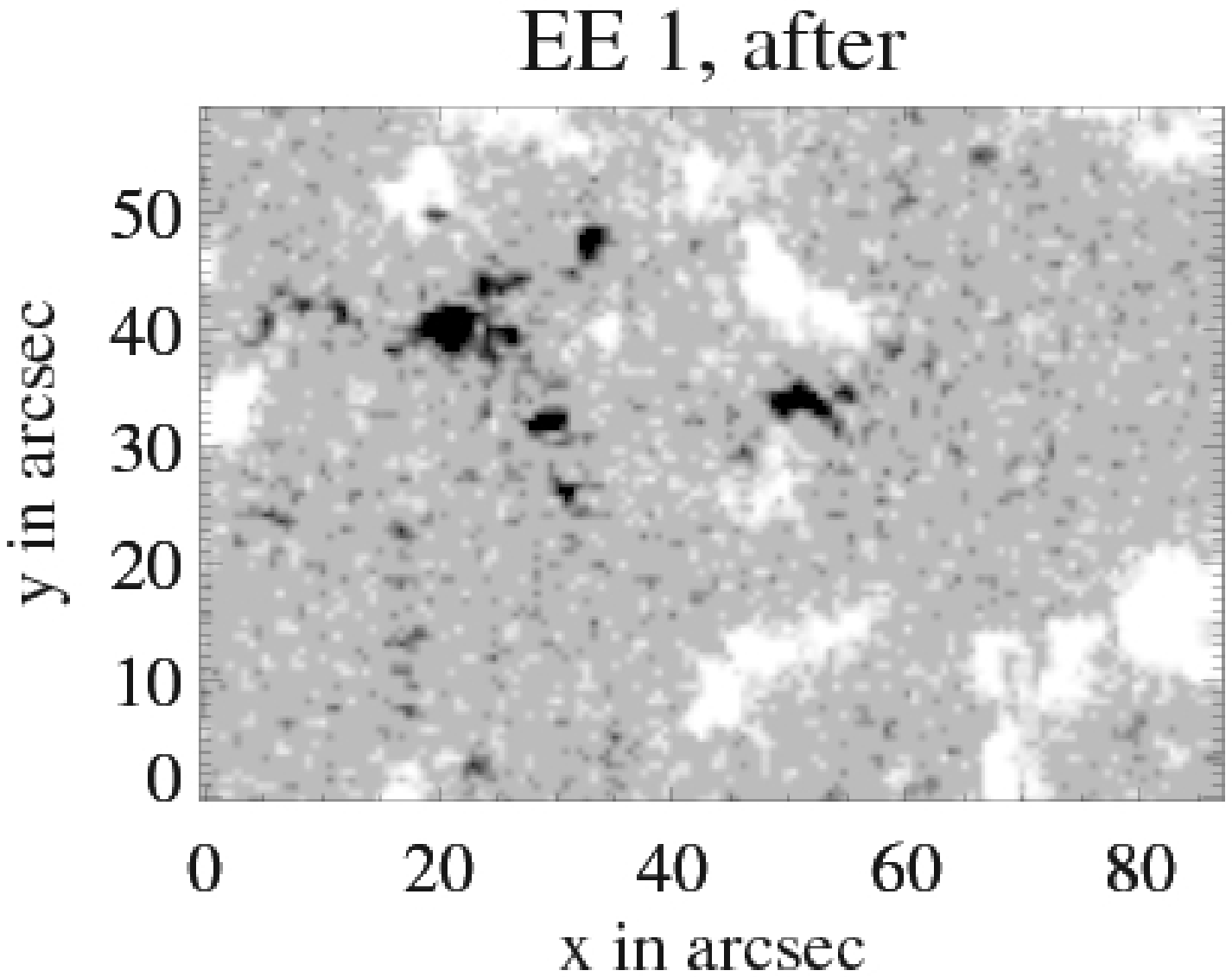}
 }
\vspace{-0.3cm}
\hbox{
       %\hspace{-3cm}
    \includegraphics[width=0.35\textwidth]{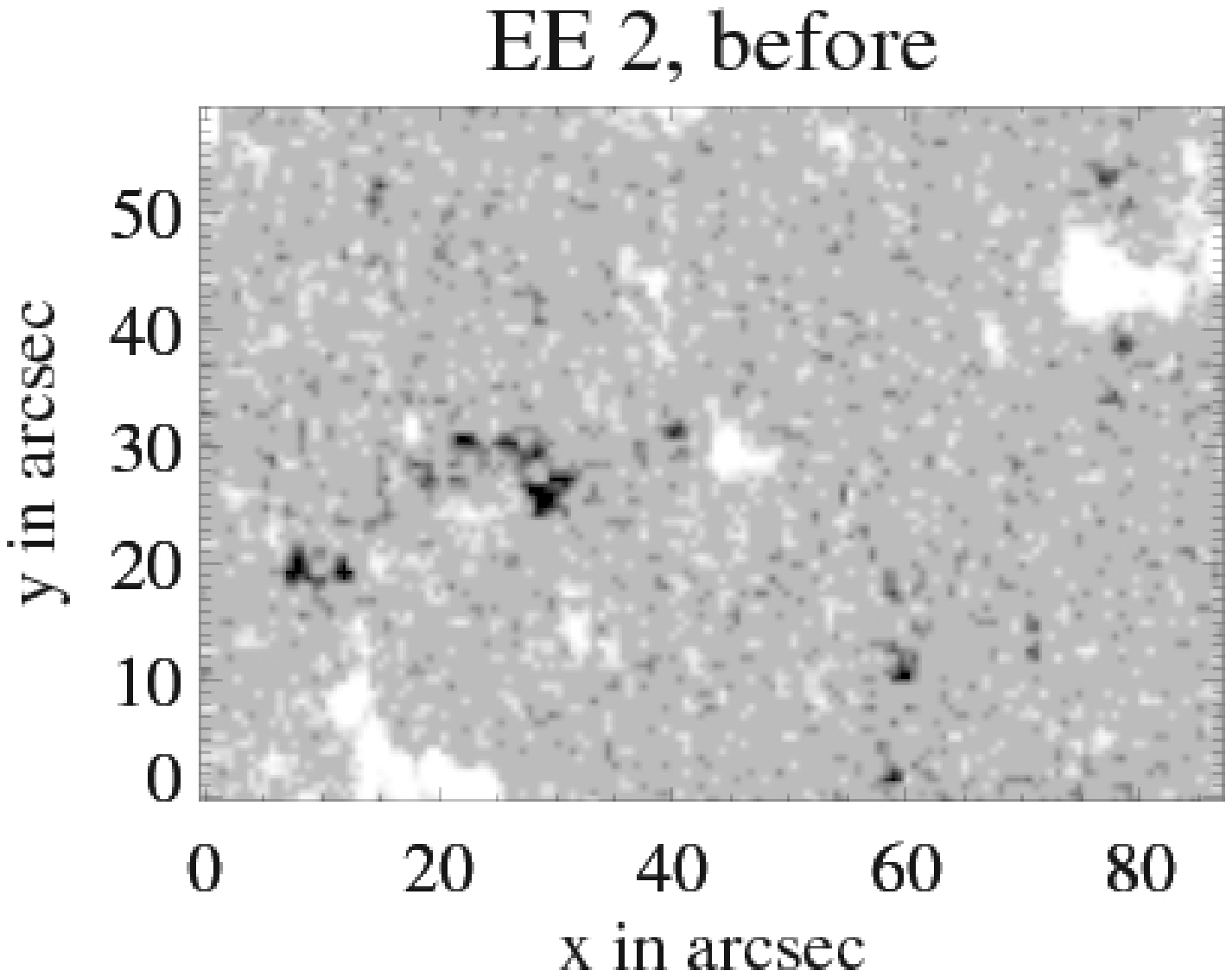}
       \hspace{-1cm}
    \includegraphics[width=0.35\textwidth]{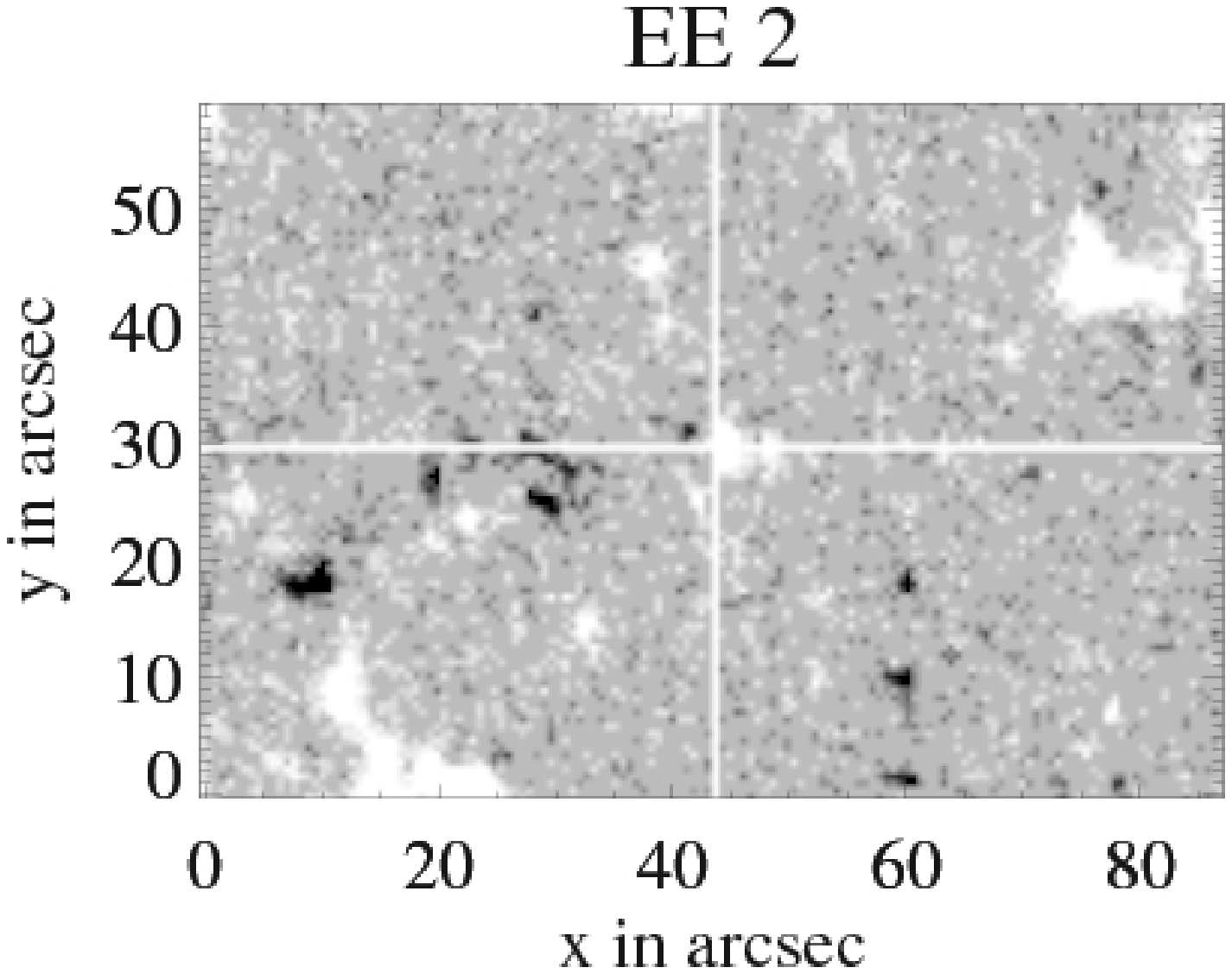}
       \hspace{-1cm}
    \includegraphics[width=0.35\textwidth]{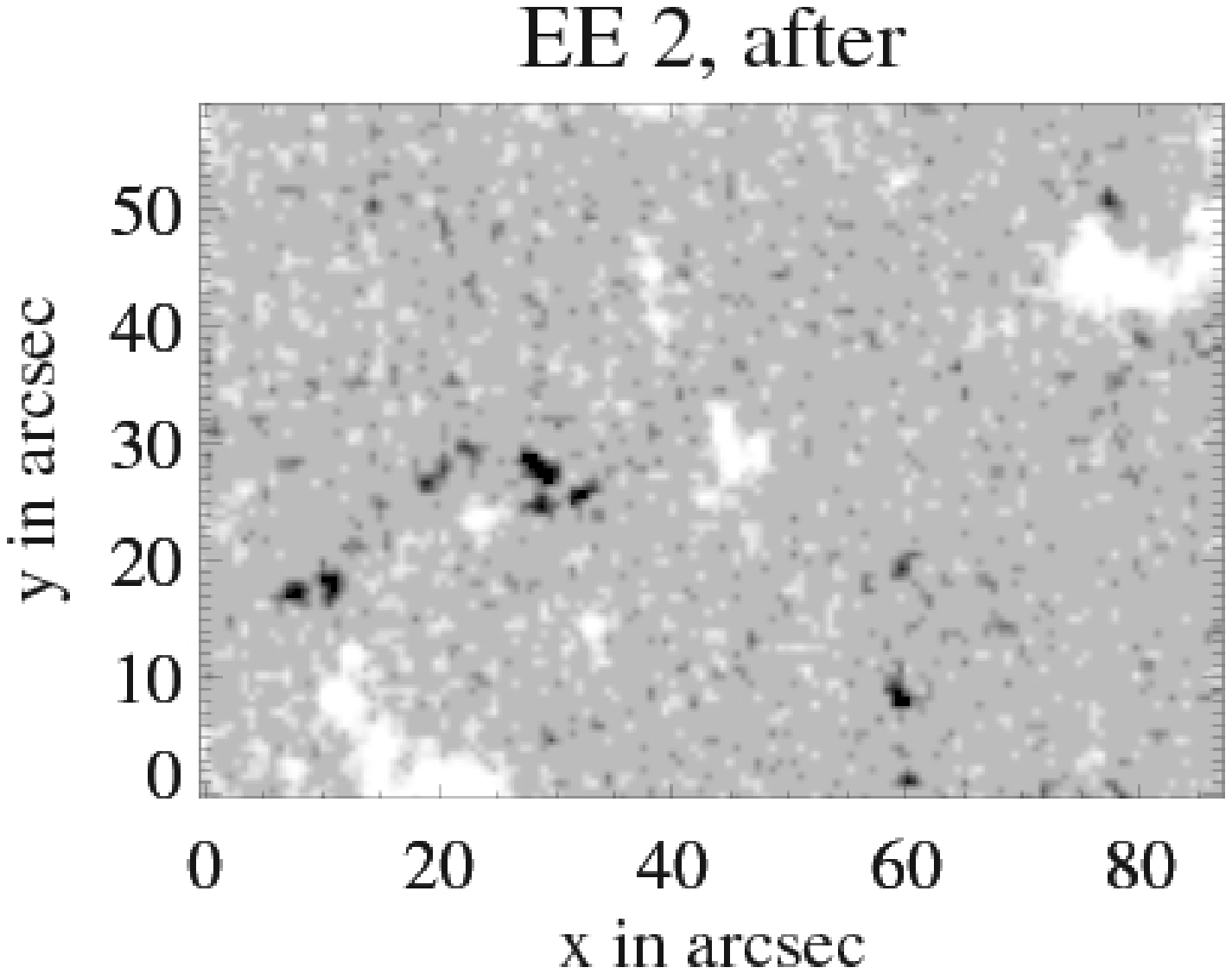}
 }
\vspace{-0.3cm}
\hbox{
       %\hspace{-3cm}
    \includegraphics[width=0.35\textwidth]{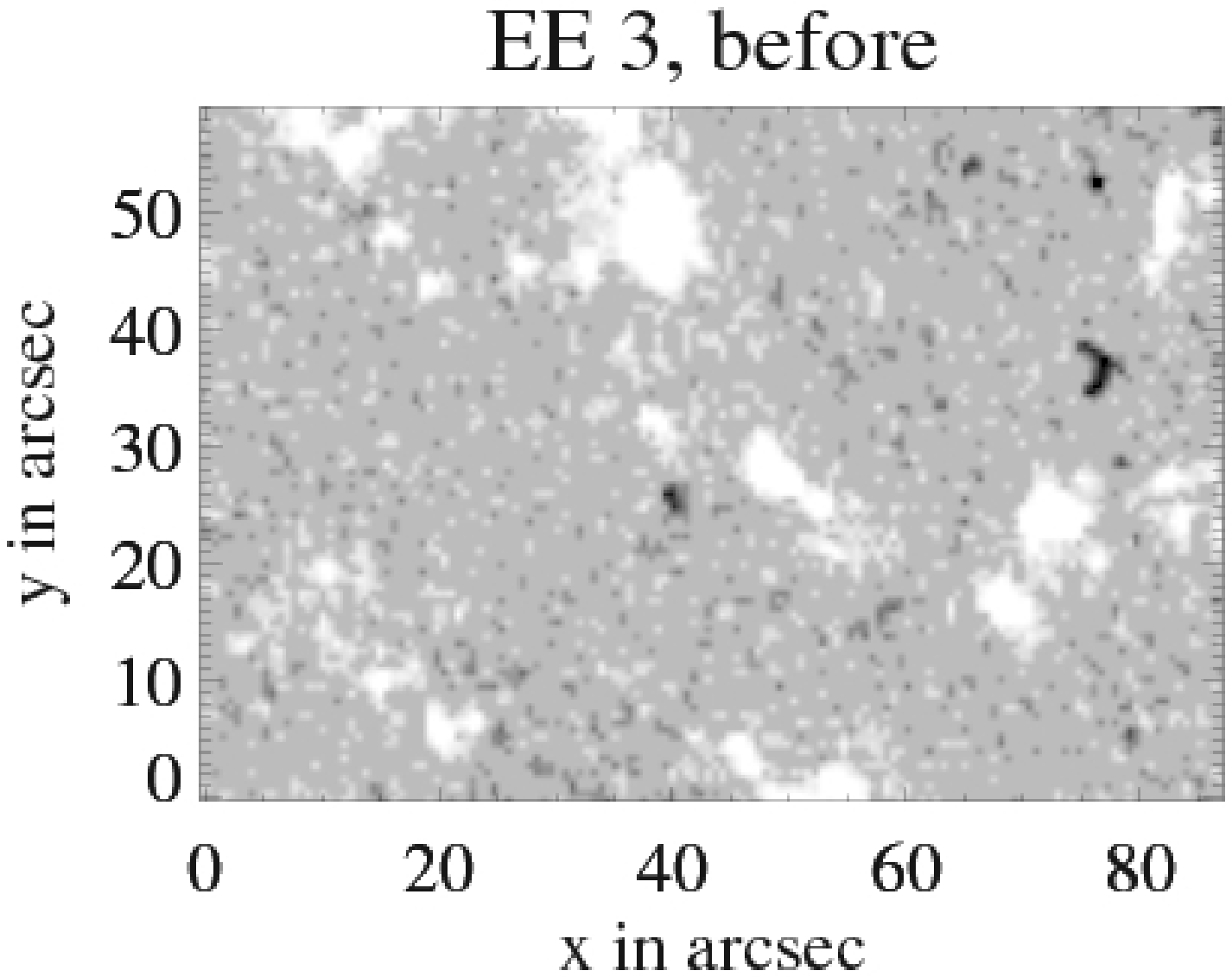}
       \hspace{-1cm}
    \includegraphics[width=0.35\textwidth]{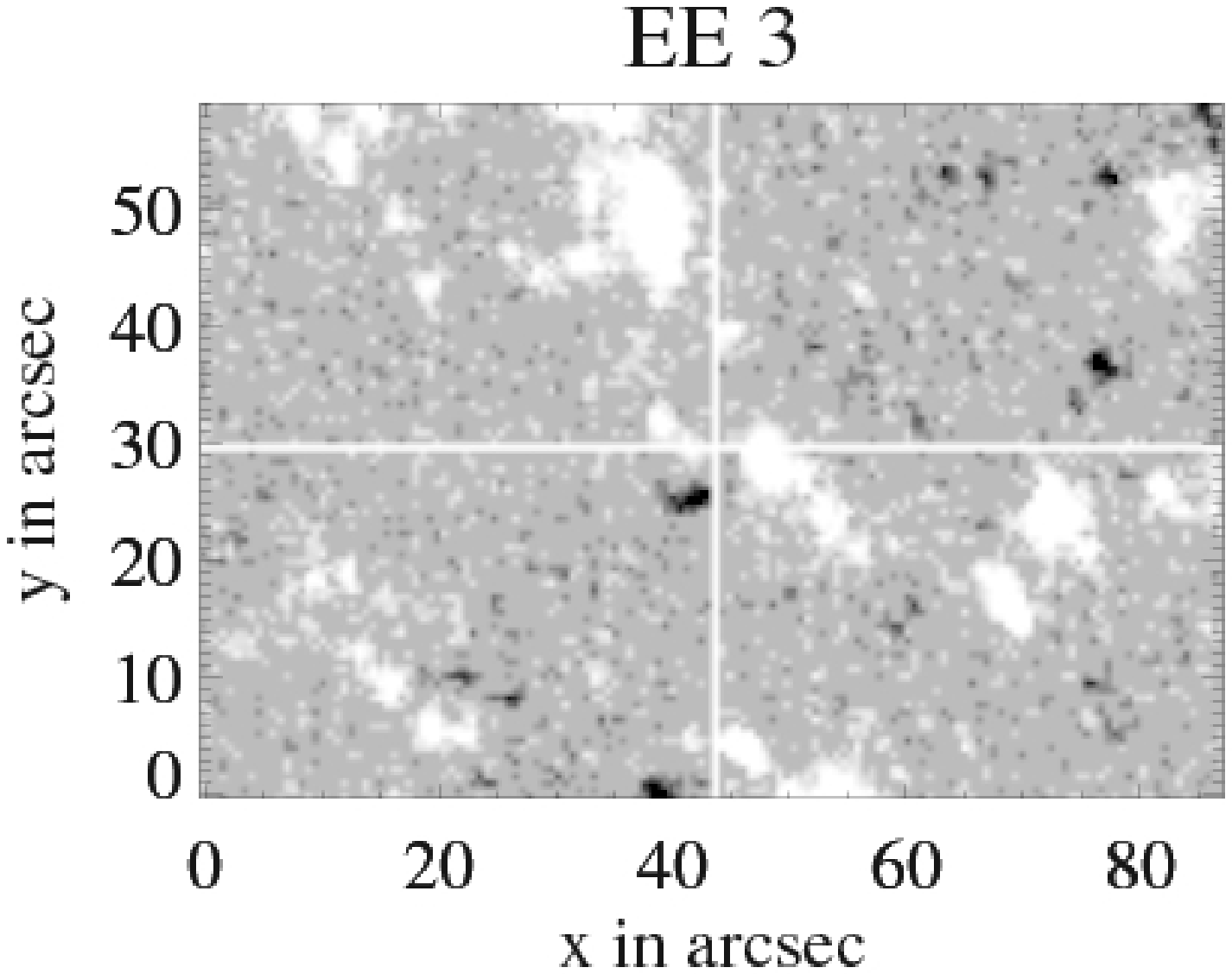}
       \hspace{-1cm}
    \includegraphics[width=0.35\textwidth]{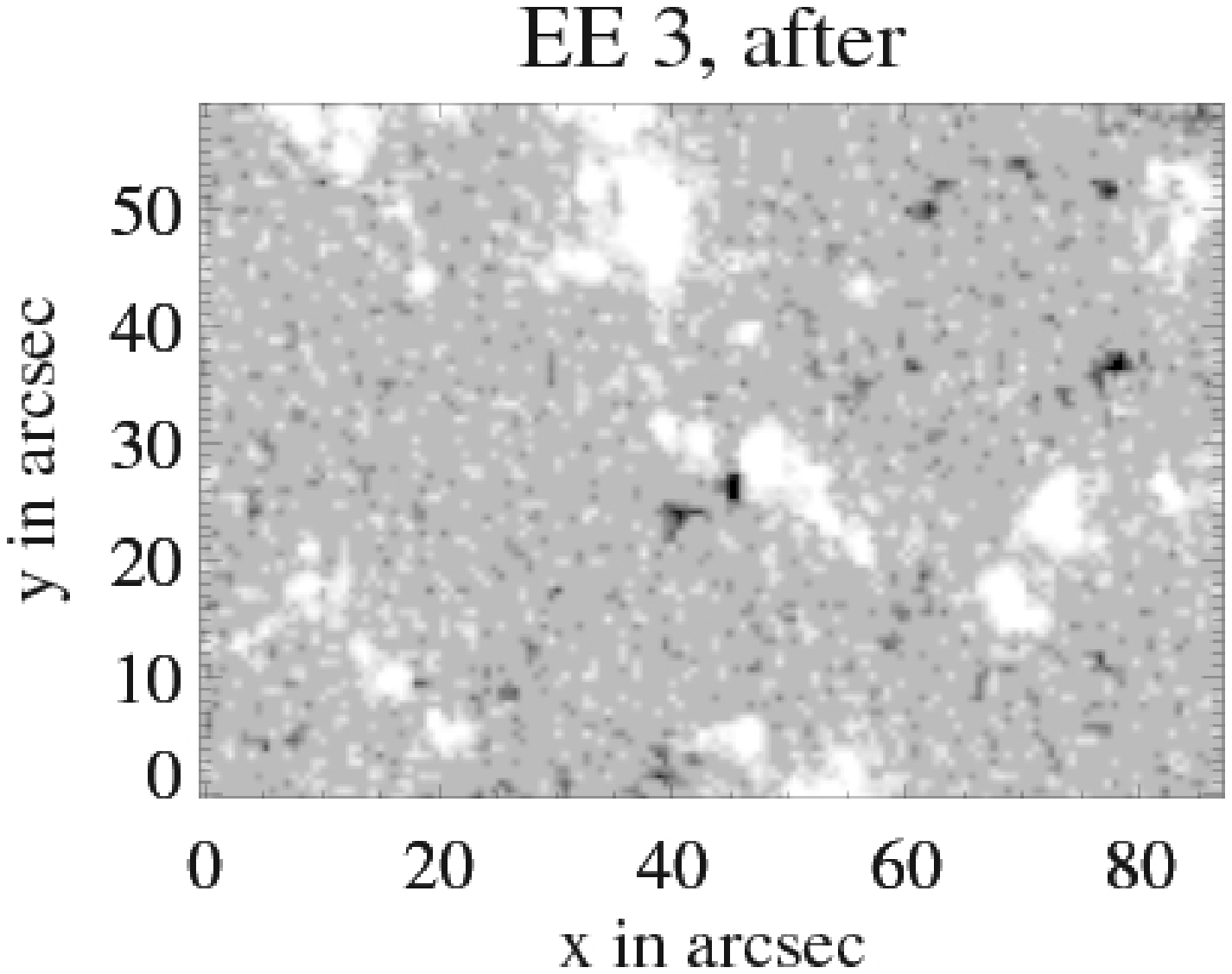}
 }
\vspace{-0.3cm}
\hbox{
       %\hspace{-3cm}
    \includegraphics[width=0.35\textwidth]{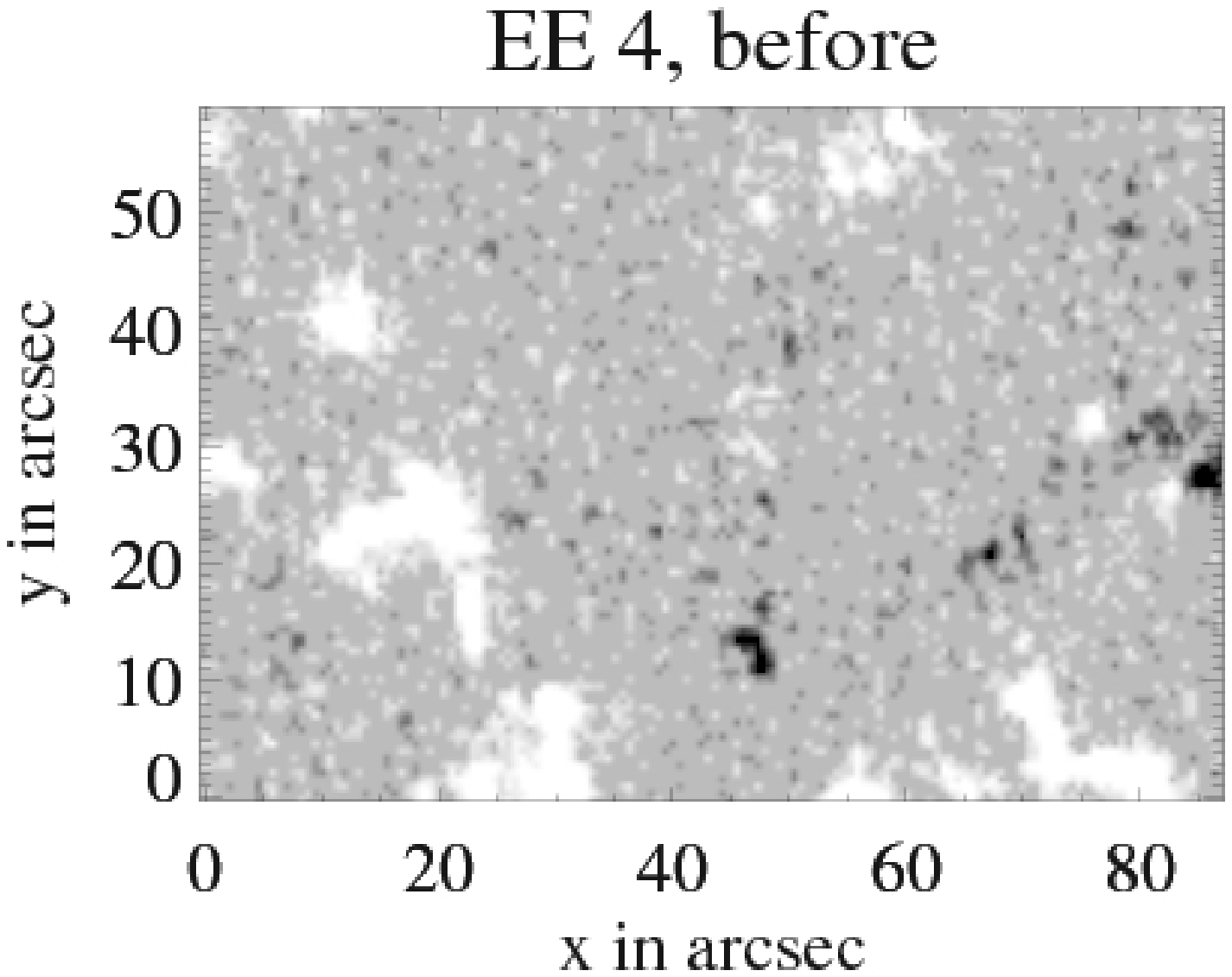}
       \hspace{-1cm}
    \includegraphics[width=0.35\textwidth]{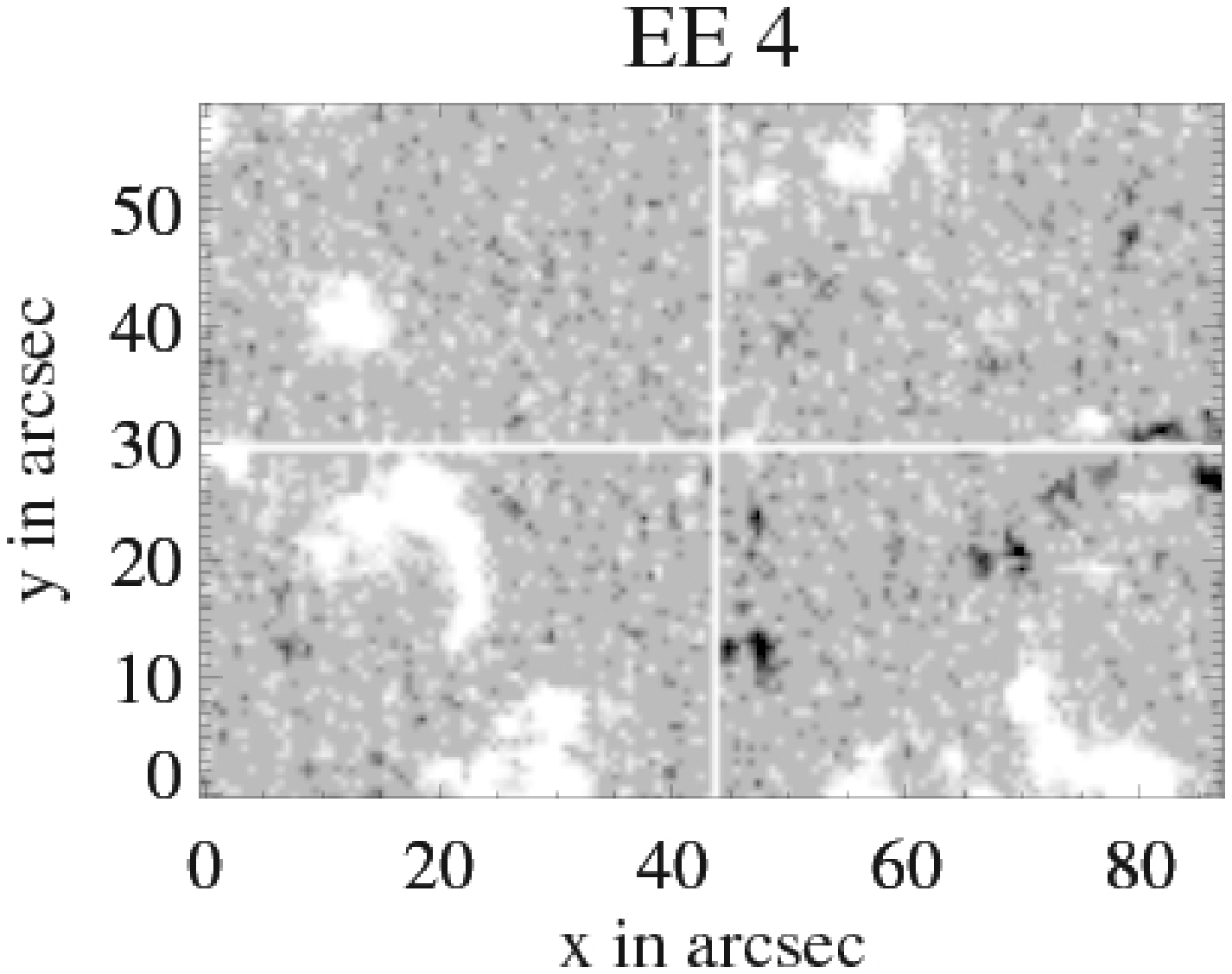}
       \hspace{-1cm}
    \includegraphics[width=0.35\textwidth]{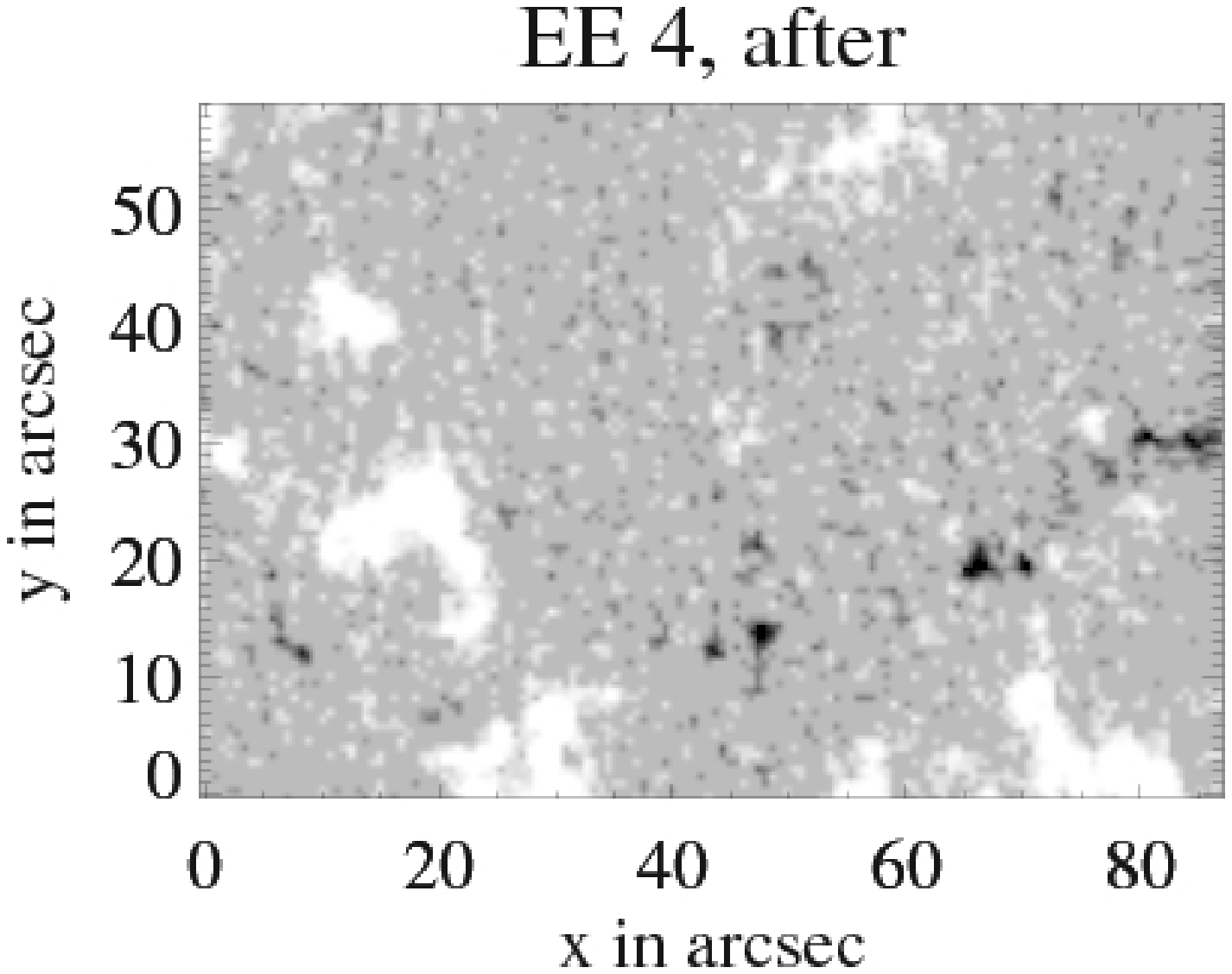}
 }
\vspace{-0.3cm}
\hbox{
       %\hspace{-3cm}
    \includegraphics[width=0.35\textwidth]{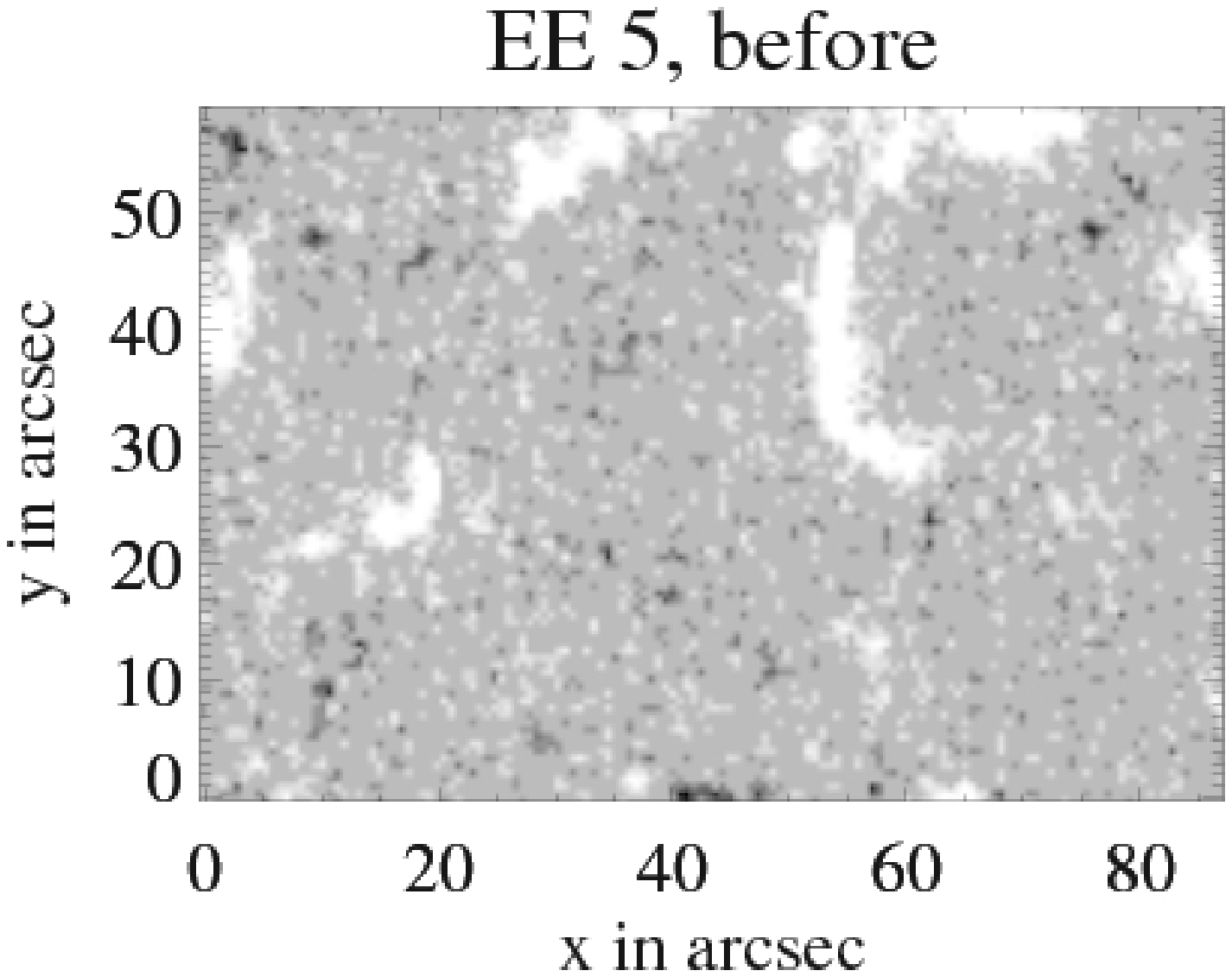}
       \hspace{-1cm}
    \includegraphics[width=0.35\textwidth]{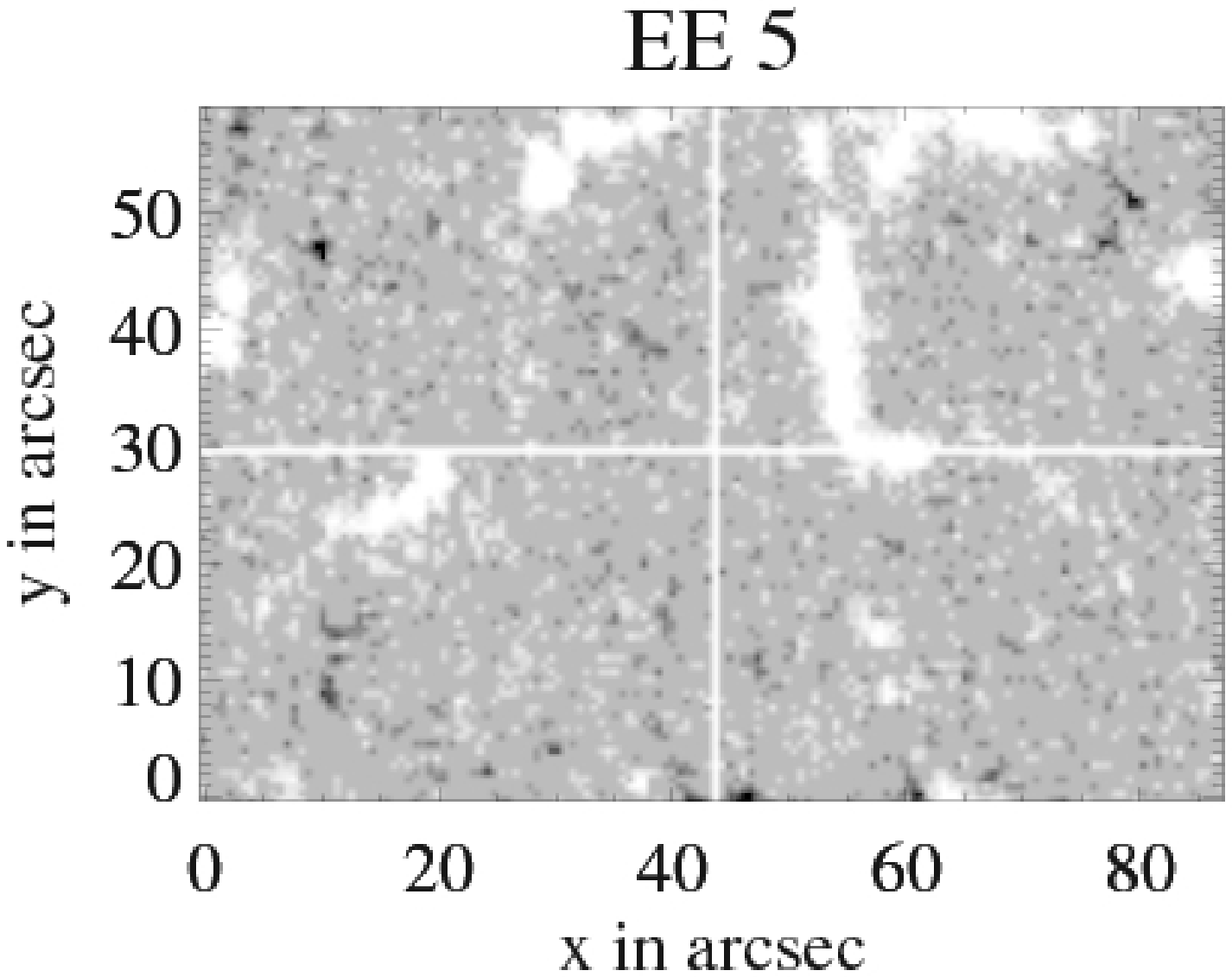}
       \hspace{-1cm}
    \includegraphics[width=0.35\textwidth]{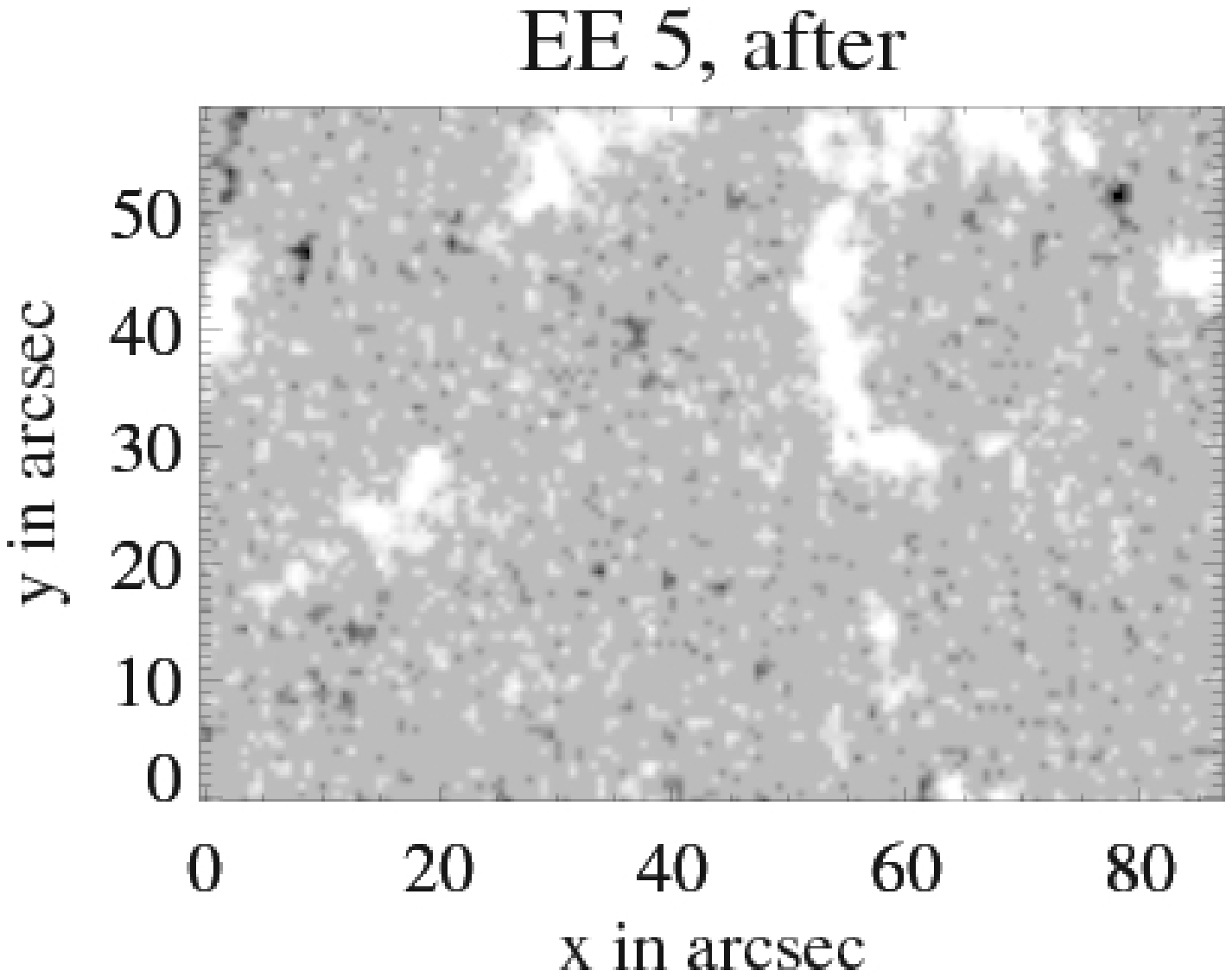}
 }
}
\caption{
   Examples of magnetic flux evolution at the location of
   an EE, images are saturated at $B = \pm$ 30 Mx cm$^{-2}$.
   Left: a magnetogram taken 30 min before the
   EE took place. Center: a magnetogram at the time of the
   EE, with the EE being located in the center of the FOV
   (indicated by the white cross). Right: a magnetogram
   taken 30 min after the EE. The 2 top rows show flux cancellation,
   the third row flux emergence, the fourth row complex flux
   changes and in the bottom row there are no flux changes.
        }
\label{fig:MDI-evol}
\end{figure}

All of our SUMER observations were taken
in very quiet regions. Comparing the observed
FOV with EIT EUV images we find that the observations
were taken in a coronal hole. Therefore,
the observed MDI flux densities are very
low. As can be seen from Fig.~\ref{fig:SUMER-MDI-scan},
EEs are usually located near, but not directly inside the
bright network (although a couple of exceptions can be found).
While the flux density of the network is
between 100 \mxcm\ and several hundred \mxcm ,
values of the cancelling/emerging flux
are between 20 \mxcm\ and less than 100 \mxcm .
%(except when the minority flux is cancelling with
%a majority flux that is part of the network).
The sizes of the flux patches are also very small
(a few arcsec): usually just small bipoles or one small
unipolar patch of flux cancelling with a larger network
fragment. The duration of the flux changes cannot be
determined as our 1h time window generally includes
only part of the process. Because we only see snapshots of
EEs in our SUMER data, such an investigation would
require continuous observations at a given location.

From Table \ref{tbl-1} we can see that the number
of EE sites with flux changes clearly exceeds the number
of random locations that show flux changes.
Therefore, we believe that the flux changes we find
%Therefore, we can infer, that the flux changes we find,
are not just accidental occurrences.
The ratio of flux cancellation to flux emergence
is 7:3. This is very close to the ratio of about 2:1 found by
Webb et al.~(1993) studying the correspondence between
X--ray bright points and evolving magnetic features in the
quiet Sun, and Harvey (1985) for He I dark points.
As explained in our introduction X--ray bright
points emit at higher temperatures, are larger and have longer
lifetimes than EEs. The flux involved is also stronger than
what we measure in EEs. Nevertheless, both studies give about
the same ratio. Webb et al. also find several cases of both
emerging and cancelling flux, corresponding to what
we call complex. In summary, we find several similarities
between our results and the results of Webb et al. with
regard to their magnetic field data, although we do not
consider EEs smaller versions of X--ray bright points.
Webb et al.~(1993) also found that there
were many more flux cancellation sites than
X--ray bright points. We can speculate that one of the
reasons might be that, to produce X--ray bright points
reconnection has to occur in the corona, while
reconnection taking place in the TR will produce EEs.
In case the reconnection happens in even lower layers,
e.g., in the photosphere, we would still measure flux
cancellation, but neither EEs nor X--ray bright points.

To properly interpret the MDI magnetograms we need
to understand how the flux densities are derived.
The quantity that an imaging magnetograph measures
is the circular polarization or Stokes $V$.
Based on the longitudinal Zeeman effect this Stokes $V$
signal is interpreted as originating from a magnetic field
permeating the plasma that the spectral lines originates
from (e.g., Stenflo 1994, Lites 2000).
In a filtergraph Stokes $V$ is usually measured in the
wing of a spectral line, where the polarization signal
is largest. In case of MDI the polarization is measured
at several wing locations (Scherrer et al.~1995).
A number of assumptions are made in converting the resulting
signal to a magnetic field.
Only the component of the magnetic field vector ${\bf B}$ along
the line--of--sight produces circular polarization
(the component in the plane perpendicular to the line--of--sight
produces linear polarization, Stokes $Q$ and $U$, which are
not measured).
If one observes at disk center and the direction of the
magnetic field vector is radial, then the LOS
component can be equal to the actual field strength.
Whenever the inclination angle between the
field vector ${\bf B}$ and the line--of--sight ($\gamma$) 
changes however, so will the measured polarization signal.
%The inclination angle can be derived from the measurement of
%the linear polarization, but Stokes $Q$ and $U$ are not
%available with MDI and it is therefore unknown.

We would like to add that the assumption that the polarization
signal is proportional to the magnetic field strength is valid only
in the so--called weak field regime, that is when, the Zeeman
splitting is small compared to the Doppler width of the spectral line.
This regime depends on the spectral line and usually breaks down
at high field strengths like in a sunspot.

Another parameter that influences Stokes $V$ is the
magnetic filling factor (FF). If one assumes a simple two--component
model of the magnetic field, then the pixel under consideration
contains contributions from both a magnetized and an unmagnetized
atmosphere. In the conversion of the MDI polarization signal
it is assumed that the observed pixel is completely permeated
with magnetic field of constant strength (FF=1).
But a field that is e.g.~twice as strong but only covers
half the area of the spatial resolution element will give
the same polarization signal.
Since the filling factor cannot be determined with filtergrams,
the quantity that MDI delivers is a line--of--sight
magnetic flux density, not a field strength $|{\bf B}|$.
From full Stokes polarimetry we know that only sunspot
umbrae have FF$\approx$1.
For the quiet Sun, one has to differentiate between the magnetic
network and internetwork.
Network (and plage regions) are known to consist of
magnetic elements with field strengths of about 1500 G and
sizes of around 100 km ($<$ 0.2\arcs ).
Filling factors for these regions range from 
FF=0.05 in quiet network up to 0.24--0.40 in plage regions
(e.g.~Stenflo 1973, Muglach \&\ Solanki 1992, Mart\'\i nez Pillet
et al.~1997, Bellot Rubio et al.~2000).

The topology of the internetwork field is controversial.
Depending on the assumed model and the spectral lines
used, fields with moderate field strength around several
dozens to hundreds of G can be found, or again kG fields
with very small filling factors (see e.g.~recent works from
Mart{\'{\i}}nez Gonz{\'a}lez et al.~2006, 2008 and
Socas--Navarro et al.~2008). New observations with the
spectro-polarimeter onboard the Hinode satellite
with higher sensitivity and no interference from
atmospheric seeing will provide new input into this
unresolved issue. First results on internetwork
fields from Hinode can be found e.g.~in Orozco Su{\'a}rez et
al.~(2007), Lites et al.~(2008) and de Wijn et al.~(2008).

The MDI movies show fluctuations of small flux densities
on very short temporal and spatial scales.
These flux densities are above $B = \pm$ 9 Mx cm$^{-2}$, which is
our one sigma noise level.
Most of these fluctuations are probably caused by changes
in inclination $\gamma$ due to the swaying of the field lines
in the granular velocity field, perhaps in addition to changes
in FF as the field lines are shuffled together by the granules.
As these fluctuations take place all over the observed field,
we think that they are unrelated to EEs, or one would
find a lot more EEs along the slit. The flux changes we
consider significant for EEs involve somewhat larger flux patches
that are persistent over more than one granular lifetime.
We think that due to these rather conservative criteria,
we do not find flux changes for the majority of our EEs.
In addition, the magnetic sensitivity and spatial resolution
of MDI are very limited. We hope that future observations from
ground and especially from space (e.g.~from Hinode/SOT) will
improve these statistics and contribute to our understanding of
small-scale photospheric dynamics reflected in magnetograms.

Visual inspection of the Dopplergram sequences did not reveal any
velocity signatures we would expect in connection with the
restructuring of the magnetic field. If flux emergence is
really connected to flux elements moving from subphotospheric
layers upward into the photosphere and higher layers,
eventually leading to reconnection with pre-existing flux in
the TR, one would expect to see upflows in connection with flux
emergence.  Or if flux submerges as a consequence of flux
cancellation, one would expect to see downflows at the location
of the cancelling flux (under the assumption that the
reconnected loop is actually pulled under the surface).
After p--mode filtering the Doppler velocity
field shows fluctuations of the order of $\pm$ 600 \ms\
due to granular and supergranular flows. Beyond these fluctuations,
however, no clear signature of e.g.~downflows can be found at flux
cancellation sites. However, the MDI Dopplergram signal is dominated
by the non-magnetic part of the atmosphere within the observed
pixel, while the magnetic field covers probably only a small
fraction of the area. A better way to investigate this
question would be to measure Stokes $V$ profiles and search for
velocity signatures in the Stokes $V$ zero-crossing
(e.g.~Muglach \& Solanki 1992) which reflects Doppler shifts
originating exclusively from the magnetic part of the
atmosphere. Increasing the spatial resolution should also
increase the likelyhood of detecting velocity signatures. 

As mentioned in the Introduction Chae et al.~(1998) also
compared SUMER EEs with line--of--sight magnetograms taken
at BBSO. They found that about 64\% of their EEs occurred during
magnetic flux cancellation, while they did not mention any
flux emergence. Their result differs considerably
from our statistics and hence deserves closer inspection.
The SUMER spectra that they analysed were taken in a
sit--and--stare mode of the spectrograph. Therefore,
due to the lack of solar rotation compensation solar structures
moved through the 1\arcs\ slit in about 420s.
During the 4.5h of their observing run, SUMER covered
a region of about 40\arcs\ from which they construct
a spectroheliogram that is compared with the magnetograms.
They explained their method of identifying EEs in the
SUMER spectra, but did not provide information on how
EEs were counted in their analysis.
%Were adjacent spectra
%along the slit or EE signatures along the slit counted as one?
Their Figure 1 shows the recurrence of an EE in a sequence
of exposures, but it is unclear whether this would be
considered to be one or two EEs in their statistics.
This lack of information complicates an actual comparison of the
respective results.
The magnetograph data in their Figure 2 suggests that there
are just two locations of flux cancellation,
but it is unclear how many EEs were actually assigned to them.
In contrast, our SUMER data represent snapshots of the TR with no
information about the evolution of the EEs, whose
lifetimes are comparable with our exposure times and
sizes similar to the step size of the scan.

Unlike Chae et al.~(1998), we also find flux emergence
and complex flux changes in connection with EEs.
Therefore, we propose that what we see is a signature of
reconnection related to a {\it change in magnetic topology}
of the small--scale magnetic field of quiet sun regions.
This is a more general concept than flux cancellation.

A number of modelling efforts have been reported,
e.g.~Karpen et al.~(1998), Sarro et al.~(1998),
Roussev \& Galsgaard (2002) and Chen \& Priest (2006).
Chen \& Priest (2006) model EEs under the influence
of 5 min p-mode oscillations, trying to reproduce
the recurrence rate of 5 min reported by Ning et al.~(2004).
They do get a 5 min modulation in their synthetic TR
line, but the modulation is present in the central component
of the line whereas no enhanced line wings are present.
Some of the simulations of Roussev \& Galsgaard (2002) on
the other hand reproduce the line wing enhancements
of their TR lines very well (see e.g.~their Figure 5), but
also produce wing enhancements in the coronal line
Ne VIII 770 \AA , which are usually not observed (most
of the EEs in our sample are observed in C IV 1548 \AA\ 
and also include simultaneous observations of Ne VIII
770 \AA\ ). 
One of the challenges for modelling seems to be
the fact that EEs are observed in TR lines, but not
in coronal lines. The reconnection scenario therefore
has to produce appropriate acceleration of sufficient
plasma (not just heating) at TR temperatures, without heating
and accelerating the plasma to coronal temperatures.
Up to now modelling efforts are still too simplified to
allow a detailed comparison with observations. 
We hope that the results of our detailed
study of the photospheric field evolution will
provide valuable input into future models.

%__________________________________________________________________

\acknowledgments

We would like to thank the SUMER team, especially W.\,Curdt
and U.\,Schuehle and the MDI team for help with the data.
Valuable discussions and help from J.\,Cook, K.\,Dere,
J.\,Mariska, J.\,Karpen, I.\,Ugarte--Urra and Y.--M.\,Wang are
gratefully acknowledged.
This work was supported by NASA's Sun-Earth Connection
Guest Investigator program (project \# NNG04ED07P),
which is gratefully acknowledged.
SoHO is a project of international cooperation between ESA and NASA.

%__________________________________________________________________

%% ----------------------------------------------------------------------
%% --- FIGURES ----------------------------------------------------------
%% ----------------------------------------------------------------------

\clearpage

\clearpage

\clearpage

%\includegraphics[bb=100 70 410 370, width=8.0cm]{f01.eps}
%\includegraphics[width=5cm]{f02.eps}

%% ---------------------------------------------------------------------
%% --- TABLES ----------------------------------------------------------
%% ---------------------------------------------------------------------

\clearpage
\begin{table}
%\begin{center}
\caption{Magnetic flux changes observed during explosive events \label{tbl-1}}
\vspace{0.7cm}
\begin{tabular}{ccccccc}
\tableline\tableline
 & & flux changes   & &  & no flux changes &  total number \\
\tableline
  & &  &  & all flux &  & \\
 & emergence &  cancellation & complex & changes & &  \\
 &  &  &  & & & \\
\tableline
 &  &  &  & & & \\
explosive events & 3 & 7 & 4 & 14/38\% & 23/62\% & 37 \\
random locations & 2 & 1 & 1 &  4/11\% & 33/89\% & 37 \\
 &  &  &  & & & \\
\tableline\tableline
\end{tabular}
%\end{center}
\end{table}

\end{document}